\documentclass[11pt,a4paper]{article}
\pdfoutput=1
\usepackage[pdftex]{graphicx}
\usepackage{jheppub}
\usepackage{bm}
\usepackage{amsmath} 
\usepackage{amsthm}
\usepackage{amssymb} 
\usepackage{MnSymbol}
\usepackage{hyperref}
\usepackage{xfrac}
\usepackage{subfig}
\usepackage{booktabs}

\usepackage[inline]{enumitem}

\newcommand{\be}{\begin{equation}}
\newcommand{\ee}{\end{equation}}
\newcommand{\bea}{\begin{eqnarray}}
\newcommand{\eea}{\end{eqnarray}}

\def\tr{{\mathrm{tr}}}
\def\demand{\overset{!}{=}}
\def\CH{\mathcal{H}}
\def\CK{\mathcal{K}}
\def\CL{\mathcal{L}}

\def\CN{\mathcal{N}}

\def\CR{\mathcal{R}}
\def\CW{\mathcal{W}}

\def\slashchar#1{\ensuremath{                               %
   \setbox0=\hbox{${}#1{}$}       
   \dimen0=\wd0                                 
   \setbox1=\hbox{/} \dimen1=\wd1               
   \ifdim\dimen0>\dimen1                        
      \rlap{\hbox to \dimen0{\hfil/\hfil}}      
      {}#1{}                                    
   \else                                        
      \rlap{\hbox to \dimen1{\hfil${}#1{}$\hfil}}   
   \fi}}

\begin{document}

\title{Searching for Fermi Surfaces in Super-QED}
\author[1]{Aleksey Cherman,}
\emailAdd{cherman@physics.umn.edu}
\author[2]{Sa\v{s}o Grozdanov}
\emailAdd{s.grozdanov1@physics.ox.ac.uk}
\author[2]{and Edward Hardy}
\emailAdd{e.hardy12@physics.ox.ac.uk}
\affiliation[1]{Fine Theoretical Physics Institute, Department of Physics, University of Minnesota, \\
116 Church St SE, Minneapolis MN 55455 USA}
\affiliation[2]{Rudolf Peierls Centre for Theoretical Physics, University of Oxford, \\ 1 Keble Road, Oxford OX1 3NP, United Kingdom}

\date{\today}

\abstract{The exploration of strongly-interacting finite-density states of matter has been a major recent application of gauge-gravity duality. When the theories involved have a known Lagrangian description,  they are typically deformations of large $N$ supersymmetric gauge theories, which are unusual from a condensed-matter point of view.  In order to better interpret the strong-coupling results from holography, an understanding of the weak-coupling behavior of such gauge theories would be useful for comparison. We take a first step in this direction by studying several simple supersymmetric and non-supersymmetric toy model gauge theories at zero temperature. Our supersymmetric examples are $\mathcal{N}=1$ super-QED and $\mathcal{N}=2$ super-QED, with finite densities of electron number and R-charge respectively.   Despite the fact that fermionic fields couple to the chemical potentials we introduce, the structure of the interaction terms is such that in both of the supersymmetric cases the fermions do not develop a Fermi surface.  One might suspect that all of the charge in such theories would be stored in the scalar condensates, but we show that this is not necessarily the case by giving an example of a theory without a Fermi surface where the fermions still manage to contribute to the charge density.}


\begin{flushright}OUTP-13-13P\\
FTPI-MINN-13/27, UMN-TH-3217/13 \end{flushright}
\maketitle
\flushbottom

\section{Introduction}
\label{sec:Introduction}
Understanding the behavior of quantum matter at finite temperature $T$ and density $\mu$ is a major challenge in many areas of physics, ranging from traditional condensed matter topics to quark-gluon plasmas as explored at RHIC and the LHC, to the behavior of super-dense QCD matter in the cores of neutron stars. Developing such an understanding is  especially difficult when the systems are strongly coupled and traditional perturbative techniques are not useful.  One powerful non-perturbative technique which has attracted a great deal of attention in recent years is gauge-gravity duality \cite{Maldacena:1997re}, which maps questions about some special strongly-coupled field theories to questions about weakly-coupled theories of gravity, which are much easier to work with.  (For reviews, see e.g.~\cite{Hartnoll:2009sz,McGreevy:2009xe}, and especially \cite{Iqbal:2011ae}, which has a focus on Fermi surfaces.) This has led to many interesting results for the study of finite-density quantum matter, but also a number of puzzles, such as the fate of Fermi surfaces in the strongly-coupled systems which have gravity duals.

The ability to do controlled calculations on the gravity side of the duality comes with several conditions and costs.  To justify treating the gravity side of the duality classically, which is in general the only tractable limit, one needs the field theory to be (1) strongly coupled, typically in the sense of having a tunable 't Hooft coupling which is taken to be large and (2) to be in some kind of large $N$ limit.  Indeed, in all of the cases where the dual field theory Lagrangian is explicitly known, the field theory is a non-Abelian gauge theory, and the parameter $N$ is associated with the rank of the gauge group.\footnote{Finding such a large parameter in the known phenomenologically-relevant examples is a challenge, especially in the examples from condensed matter.}  Finally, the class of theories which have strong-coupling limits and a large $N$ limit is clearly rather special,\footnote{For instance, large $N$ QCD is not such a theory, since its 't Hooft coupling runs, and is thus not a tunable control parameter.} and in all of the cases where the dual field theory Lagrangians are known, they are supersymmetric gauge theories or deformations thereof, see {\it e.g.} \cite{Maldacena:1997re,Karch:2002sh,Gauntlett:2009dn,Gauntlett:2009bh,Gubser:2009qm} for some prototypical examples.  

These considerations make it difficult to tell a priori which of the many interesting results gauge-gravity duality has yielded are due to strong coupling, large $N$, the special nature of the field content and interactions in the theories which have gravity duals, or some combination of these.  In this sense, gauge-gravity duality is essentially a black box, since it is only tractable in a limit where the field theory description is fundamentally difficult to work with.  Moreover, while the duality has yielded many striking results, it has also produced many mysteries, such as the fate of Fermi surfaces at strong coupling, explored in e.g. \cite{Lee:2008xf,Liu:2009dm,Edalati:2010hk,Edalati:2010pn,Hartnoll:2010gu,Hartnoll:2011dm,Gauntlett:2011wm,Gauntlett:2011mf,Belliard:2011qq,Huijse:2011hp,Huijse:2011ef,Sachdev:2011ze,Ogawa:2011bz,Hartnoll:2012wm,Faulkner:2012gt,Anantua:2012nj,Sachdev:2012tj,DeWolfe:2011aa,DeWolfe:2012uv}.  The `microscopic' field content of the theories with gravity duals generally includes gauge bosons, fermions, and scalars, with the number of degrees of freedom for all of these scaling as $\mathcal{O}(N^2)$ in the 4D field theory examples.  In these theories chemical potentials for conserved charges usually couple to \emph{both} the scalars and fermions at the microscopic level.  Hence if intuition derived from studies of weakly-coupled non-supersymmetric theories were to be boldly applied to the strong coupling limit of the kind of theories which have gravity duals, then one might have expected that Fermi surfaces would be ubiquitous in systems with gravity duals.  

However, while Fermi surfaces have shown up in some examples of gauge-gravity duality, they do not seem to be at all ubiquitous.   Signs of Fermi surfaces for the $\mathcal{O}(N^2)$ degrees of freedom have recently shown up in e.g. \cite{Hartnoll:2011dm} in correlation functions of fermionic operators in electron star geometries \cite{Hartnoll:2009ns,Hartnoll:2010gu}, and in some top-down calculations in \cite{DeWolfe:2011aa,DeWolfe:2012uv} for 4D theories.   But in other examples Fermi surfaces appear to be absent \cite{Gauntlett:2011mf,Gauntlett:2011wm}.  Meanwhile, Fermi surfaces have been observed  in fermionic correlators of $\mathcal{O}(N^0)$ densities of probe fermions in work initiated in \cite{Liu:2009dm,Lee:2008xf}.  To make the situation more complicated, naively --- that is, based on expectations from weak-coupling studies of systems familiar from condensed matter --- Fermi surfaces should have an imprint on bosonic correlation functions as well, showing up as e.g. momentum-space singularities in density-density correlation functions leading to Friedel oscillations. Indeed, in holography one only has access to gauge-invariant observables, while Fermi surfaces for the quarks in a gauge theory would not be gauge-invariant.  So such Fermi surfaces might be `hidden' \cite{Huijse:2011ef} in the gravity duals,  and hence singularities in gauge-invariant charge density correlation functions may  seem to be especially promising places to look for traces of Fermi surface physics.  But such density-correlator signatures of underlying Fermi surfaces have not  been seen in many holographic systems.\footnote{In \cite{Anantua:2012nj} it is observed that density-density correlation functions in theories with dual Lifshitz geometries \cite{Koroteev:2007yp,Kachru:2008yh} with $z=\infty$  have momentum-space singularities which suggest the presence of a Fermi surface, but $z< \infty$ examples do not.  }

These considerations motivate our belief that to better understand the results of gauge/gravity duality calculations, it would be very useful to reexamine some observables for which strong coupling results from holography are available at weak coupling using conventional field-theory techniques, where one can see all of the moving pieces.  In particular, one would have direct access to any `hidden' Fermi surfaces, since at weak coupling it makes sense to work in a gauge-fixed formalism.  We will focus on $D = 3+1$ dimensional theories for simplicity, and confine our attention to the $T=0$ limit.  Our metric signature convention is $(+---)$.

An example of the kind of theory one might want to study at weak coupling is $\mathcal{N}=4$ super-Yang-Mills theory with a chemical potential for R-charge, where the number of charged degrees of freedom scales as $N^2$, originally studied in \cite{Gubser:1998jb,Cai:1998ji,Chamblin:1999tk,Cvetic:1999ne}. Another example, where the number of charged degrees of freedom scales as $N^1$,  is the $\CN=\,2$ gauge theory dual to $N_f$ D7 branes intersecting $N_c$ D3 branes in the `quenched' $N_f/N_c \ll 1$ approximation \cite{Karch:2002sh}.  The study of this latter flavored $\mathcal{N}=2$ system at finite quark number density was initiated in \cite{Kobayashi:2006sb}.    Calculations using the gravity side of the duality predict unusual thermodynamical features for this theory which are not known to arise from any weakly-coupled theory, with {\it e.g.}  a specific heat with the temperature scaling $c_V \sim T^6$~\cite{Karch:2008fa}, in contrast to what one might expect from a Fermi liquid where $c_V \sim T$.   Moreover, \cite{Karch:2008fa} found a gapless quasiparticle mode in the system which was argued to be Landau's zero sound mode (see also \cite{Kulaxizi:2008kv,HoyosBadajoz:2010kd,Ammon:2011hz,Davison:2011ek,Ammon:2012mu} for some further exploration of this identification).  But the $c_V$ scaling shows that the system is clearly a non-Fermi liquid, and to the extent that the dual field theory is a gauge theory with gapless gauge interactions, a zero sound mode would be surprising, at least at weak coupling, as we discuss further in Section~\ref{sec:Challenges}.  What is the origin of the curious thermodynamic properties of this system and what is the true identity of the quasiparticles modes?  It is possible that the puzzling thermodynamics is driven by some intrinsically strongly-coupled physics, or --- as explored recently in {\it e.g.}  \cite{Chang:2012ek,Ammon:2012mu,Bigazzi:2013jqa} --- the calculations of \cite{Karch:2008fa} were done in some metastable vacuum.  Another possibility, which can be explored using weak-coupling techniques, is that at least some of these properties are a consequence of the unusual field content and interactions of the field theory.   

However, as with the other theories with known field theory Lagrangians and gravity duals, the $\mathcal{N}=4$ super-Yang-Mills field theory examined in \cite{Maldacena:1997re} is quite complicated, as are its cousins discussed in the many follow-up works, and we will not address field theories with gravity duals directly in this work.  Instead, as a first step we will study a few simpler toy-model supersymmetric gauge theories. Specifically, we will explore the behavior of $\mathcal{N}=1$ super-QED (sQED) and $\mathcal{N}=2$ sQED in the presence of chemical potentials at zero temperature. Even these simple toy models show some curious features, since from a condensed-matter point of view they have unusual field content and interactions, with the chemical potential coupling to both scalar fields and fermions, which are in turn coupled to each other by the demands of supersymmetry.  

Perhaps the simplest questions one can ask about such systems concern the nature of their ground states.  Do the bosons condense, and do the fermions develop a Fermi surface?\footnote{We are very grateful to David Tong for an early discussion challenging our naive assumption that Fermi surfaces are inevitable at weak coupling.}  It seems natural to expect weakly-coupled scalars to condense at $T=0$ in response to a chemical potential, and we find that this is indeed what happens in our examples. One might expect Fermi surfaces to be a generic consequence of turning on chemical potentials that couple to weakly-interacting fermions based on a naive application of the standard Landau Fermi liquid picture, and intuitions derived from thinking about non-supersymmetric electron plasmas.  But we find that dense plasmas based on $\CN=\,1$ and $\CN=\,2$ sQED fail to be Fermi liquids in a fairly dramatic way, already at weak coupling.  While the chemical potential couples to the fermions in all of our examples, it does not lead to a Fermi surface in most of them.  This suggests another possible reason for the mysterious cases of missing Fermi surfaces encountered in holographic studies, aside from strong coupling. 

This paper is organized as follows.  First, in Section \ref{sec:Challenges}, we give an overview of our toy models, explain their unusual features from a condensed-matter perspective, and discuss what one might expect for their behavior at finite density.  The impatient reader may wish to look only at the summary in Section~\ref{sec:ExpectationsSummary}.  In Section~\ref{sec:N1QED}, we explore $\mathcal{N}=1$ sQED at finite electron number density.  Then in Section~\ref{sec:N2QED} we discuss $\mathcal{N}=2$ sQED with a finite electron number density, where we are forced to introduce some soft SUSY-breaking terms to stabilize the scalar sector.  Next, in Section~\ref{sec:Rcharge} we look at $\mathcal{N}=2$ sQED with a finite R-charge density.  Algebraically, the $\CN=\,2$ R-charged theory and its SUSY-broken cousins are our cleanest examples, and we evaluate the fermion contribution to the charge density for some examples in this class of theories. The somewhat surprising result of this investigation is described in Section~\ref{sec:RChargeFermions}.  Finally, in Section~\ref{sec:Discussion} we summarize our findings and sketch some of the many possible directions for future work.

We also make a brief comment on the existing literature on supersymmetric gauge theories at finite density using field-theoretic techniques.  The works closest in spirit to ours that we are aware of are \cite{Yamada:2006rx,Paik:2009iz} and \cite{Huijse:2011hp}.   Ref. \cite{Yamada:2006rx} studied $\CN=4$ SYM theory with R-charge chemical potentials compactified on a 3-sphere, with a focus mostly on the high-T limit, while \cite{Paik:2009iz} studied the finite-T properties of $\CN=\,2$ super-Yang-Mills (SYM) theory.  Ref. \cite{Huijse:2011ef} studied physics related to Fermi surfaces in non-supersymmetric theories inspired by 4D $\CN=4$ SYM, among other examples, but with their choice of models they did not run into many of the issues we deal with here.    We also note the important work \cite{2005PhRvB..72b4534P} exploring the interplay between Luttinger's theorem, Fermi surfaces, and Bose-Einstein condensation in the context of cold atomic gases.

Also, the study of super-QCD at finite quark-number was initiated in \cite{Harnik:2003ke} for $\CN=\,1$ supersymmetry and in \cite{Arai:2005pk} for $\CN=\,2$ supersymmetry, with an aim of understanding color superconductivity in a supersymmetric context.   However, the issue of the existence of Fermi surfaces in supersymmetric gauge theories at finite density was not examined in these papers.  Finally, the interesting recent works \cite{Barranco:2012ye,Barranco:2013vb} constructed a supersymmetric version of `BCS theory', without dynamical gauge fields, and engineered things such that there is no scalar condensation but there are Fermi surfaces.
         
\section{What should we expect?}
\label{sec:Challenges}
The standard example of a finite-density relativistic system involving fermions and gauge fields is a QED plasma, which we now briefly describe before considering supersymmetric theories.  We do this because much of our intuition for what to expect for finite-density physics is based on experience with this non-supersymmetric system.

The Lagrangian describing an electron plasma is just that of QED, involving the electron field $\psi$ and the photon gauge field $A_{\mu}$, and is very simple:
\begin{align}
\label{eq:QED}
\mathcal{L}_{\CN =\, 0} &= -\frac{1}{4} F_{\mu\nu}F^{\mu \nu} + \bar{\psi}\left(i\slashchar{D} - m\right) \psi + A_{\mu} J^{\mu} ,
\end{align}
where $D_{\mu} = \partial_{\mu} - i\mu \delta_{\mu 0} - i g A_{\mu}$, $g$ is the gauge coupling, $\mu$ is a chemical potential which couples to the charge of the electrons, and $J^{\mu}$ encodes the effects of other matter which provides a neutralizing background, such as some ions.  

The requirement of having a neutralizing background is essential.  While the addition of the chemical potential term is a gauge-invariant deformation of the theory, it couples to a gauged charge.  If one wants a finite density of matter in the vacuum in the infinite-volume limit, with a finite free energy density, then any negative charge density carried by the electrons must be compensated by a positive charge density carried by the ions.  Otherwise  one would pay an infrared-divergent energy cost for having long-range electric fields.   This is a textbook observation for QED plasmas \cite{1971qtmp.book.....F}, and is also true for non-Abelian gauge theories like QCD at high densities.\footnote{For some seminal papers exploring this issue {\it e.g.}  see \cite{Iida:2000ha,Alford:2002kj,Steiner:2002gx}, for a review see \cite{Alford:2007xm}.}  As is explained in {\it e.g.}  Section 2 in \cite{Alford:2002kj}, neutrality must be imposed even if the gauged charge is spontaneously broken, which will be relevant for our discussion of sQED.   Otherwise a finite size chunk of the degenerate matter would have electric fields outside of it which grow in strength with its size, again causing problems with the infinite-volume limit.

Before beginning a discussion of supersymmetric plasmas, and exploring to what extent they can be thought of as Fermi liquids, it is important to note that a standard dense low-temperature electron gas described by Eq.~\eqref{eq:QED} is already \emph{not} a Fermi liquid.  The issue is the long range of the electromagnetic interactions, and the subtle nature of screening due to the degenerate electrons.  While Coulomb photons pick up a screening mass in the static (zero-frequency) limit due to medium effects, the transverse (`magnetic') photons do not get a static screening mass so long as the photons do not become Higgsed.  Consequently, the magnetic photons continue to mediate long-range interactions, and this drives the breakdown of Fermi liquid theory \cite{1973PhRvB...8.2649H,1989PhRvB..4011571R,Gan:1993zz,1995PhRvL..74.1423C}.  This leads to subtle effects such as a non-Fermi-liquid scaling of the specific heat with temperature, $c_v \sim T \ln T$, among others.  At a more pedestrian level, the non-trivial momentum and energy dependence of the Coulomb screening effects in an electron plasma are such that the residual Coulomb interaction obliterates the would-be gapless Fermi zero-sound mode present in Fermi liquids, turning it into the gapped plasmon mode of the dense electron gas as explained in {\it e.g.} Chapter 16 of the textbook \cite{1971qtmp.book.....F}.  

Given these results for non-supersymmetric gauge theories at finite density, we clearly cannot assume that the $\CN=\,1$ and $\CN=\,2$ sQED plasmas should be Fermi liquids.  Nevertheless, while non-supersymmetric degenerate plasmas are not Fermi liquids, the fermions populating the plasma still have a Fermi surface, at least before considering the standard sort of pairing (superconducting) instabilities which can lead to its breakdown.  This remains true\footnote{Since electron and quark fields are not gauge invariant, the notion of a Fermi surface is easiest to discuss in a gauge-fixed setting, and understanding its effects in gauge-invariant language requires more work.  Fortunately, at weak coupling, where our attention will be confined, the use of such gauge-fixed notions will be very useful, as it is in {\it e.g.}   the standard discussions of gauge symmetry `breaking' in the Standard Model's Higgs mechanism.} even in more exotic non-supersymmetric systems, such as degenerate quark matter, and generalizations of Eq.~\eqref{eq:QED} to include condensed dynamical scalar fields in $J^{\mu}$~\cite{Gabadadze:2009jb,Gabadadze:2009dz,Gabadadze:2008mx,Bedaque:2011hs}, or some types of Yukawa interactions \cite{Baym:1975va}.    As we will see, however, even the very existence of a Fermi surface cannot be taken for granted in the supersymmetric case.

For a final observation about non-supersymmetric plasmas, we note that having $g\ll 1$ is necessary but \emph{not}  sufficient for a QED plasma to be weakly coupled.  The reason is that Coulomb interactions are, in a sense, strong at low energies, and tend to lead to the formation of bound states --- atoms ---  if the interaction energy dominates over the characteristic momenta of the electrons and ions.  Indeed, if we define  $l \equiv [3/(4\pi n)]^{1/3}$ as the inter-electron `spacing'  and denote the Bohr radius by $a_0 \equiv 1/(\alpha m)$, then it is well-known that in an electron gas the physical expansion parameter is $r_s \equiv l/a_0$, rather than $\alpha \equiv g^2/4\pi$, and one must have $r_s \ll 1$ for calculability.   We expect that our results  in the supersymmetric examples below will be reliable in a similar high density limit, but it will be important to verify this in future work by doing higher-order calculations.   For this work, we simply assume that our number densities are large enough that we do not have to worry about the formation of supersymmetric atoms, which were studied recently in \cite{Rube:2009yc,Herzog:2009fw,Behbahani:2010xa}.  In the terminology often used in the AdS/CMT literature, our focus on high density fully ionized plasmas means that we work in the \emph{`fractionalized'} regime of super QED, as opposed to the low-density atomic gas regime, which could be thought of as \emph{`confined'}.

We now turn to a discussion of the subtleties particular to supersymmetric plasmas.  To keep the discussion streamlined, we use $\CN=\,1$ sQED as our example.  The action of $\CN=\,1$ sQED is significantly more complicated than that of QED.  In addition to $\psi$ and $A_{\mu}$, supersymmetry requires the addition of selectron fields $\phi_{+},\, \phi_{-}$, as well as the gaugino $\lambda$, along with interaction terms amongst all of these mandated by the supersymmetrization of the gauge interaction.  The resulting action is
\begin{equation}
\begin{aligned}
\label{eq:sQED}
\mathcal{L}_{\CN=\,1} &= -\frac{1}{4} F_{\mu\nu}F^{\mu \nu} + \frac{1}{2}\bar{\lambda} i\slashchar{\partial} \lambda \\
&+ \bar{\psi}\left(i\slashchar{D}^{-} - m\right) \psi + \left|D^{-}_{\mu} \phi_{-}\right|^{2}+\left|D^{+}_{\mu} \phi_{+}\right|^{2}  - \left|m \phi_{-}\right|^{2}-\left|m\phi_{+}\right|^{2}   \\
&+ \sqrt{2} i g \left( \phi_{+}^{\dagger}\bar{\psi} P_{-} \lambda - \phi_{-}^{\dagger} \bar{\lambda} P_{-} \psi - \phi_{+} \bar{\lambda} P_{+} \psi   + \phi_{-} \bar{\psi} P_{+} \lambda    \right) 
- \frac{g^{2}}{2} \left( \left|\phi_{+}\right|^{2} - \left|\phi_{-}\right|^{2} \right)^{2}\\
&+\textrm{Ions},
\end{aligned}
\end{equation}
where $\lambda$ is a Majorana fermion, $P_{\pm} = \frac{1}{2}(1 \pm \gamma_5)$, $\phi_{\pm}$ are complex scalar fields,
\begin{align}
\label{eq:CovariantDerivative}
D^{\pm}_{\mu} = \partial_{\mu} \pm i \mu \delta_{\mu 0} \pm i  g A_{\mu}
\end{align}
and the $+\textrm{Ions}$ term encodes couplings to neutralizing `ion' fields.  We assume the ion sector is supersymmetric as well, and defer writing out the relevant contributions to the action for now.  The physical motivation for assuming that the ion sector is supersymmetric is that the theories we are really interested in --- the ones with gravity duals --- usually do not include dynamical non-supersymmetric sectors.  The action describing $\CN=\,2$ matter at finite density is even more complex, and we do not write it out here;  the general comments about $\CN=\,1$ sQED below also apply to $\CN=\,2$ sQED.

Before launching a search for Fermi surfaces in $\CN=\,1$ sQED, and then $\CN=\,2$ sQED, we should emphasize a few features of Eq.~\eqref{eq:sQED} which make the analysis tricky.  First, note that before considering the `ions', there is only \emph{one} continuous symmetry in Eq.~\eqref{eq:sQED}, under which the fields have the transformation properties $\psi \to e^{-i \alpha} \psi,\, \phi_{+} \to  e^{i \alpha} \phi_{+},\, \phi_{-} \to  e^{-i \alpha} \phi_{-},\, A_{\mu} \to A_{\mu},\, \lambda \to \lambda$.  So there are no \emph{separate} fermion number or scalar number symmetries, in a striking contrast to familiar non-supersymmetric theories, even ones studied in \cite{2005PhRvB..72b4534P}.  The fields are tied together by the Yukawa interactions in such a way that only a \emph{single} $U(1)$ remains. Second, we observe that as usual, the chemical potential enters the Lagrangian as the time component of a background gauge field.  So \emph{both} the selectron and the electron fields directly experience the chemical potential.  We note that in this situation, one should interpret any expectations based on Luttinger's theorem \cite{Luttinger:1960zz} or the theory of `compressible quantum matter' \cite{Huijse:2011hp} with care, since the assumptions underlying these frameworks do not apply in general to the systems we consider once the scalar fields condense.

The issue we explore in this work concerns the response of the selectrons and electrons to the chemical potential.  Let us start by considering the behavior of the scalar fields of $\CN=\,1$ sQED.  The scalar effective potential $V_{\rm eff}$ is the sum of the classical potential
\begin{align}
V^{(0)}_{\rm eff} = (|m|^2-\mu^2)(|\phi_{+}|^2 + |\phi_{-}|^2) + \frac{g^2}{2} (|\phi_{+}|^2 - |\phi_{-}|^2)^2
\end{align}
plus quantum corrections.  Interactions with the electrons and photons will contribute new terms to the bosonic effective potential starting at one loop level. But so long as the theory is weakly coupled, and the classical potential is non-vanishing, the selectron ground state should be determined by $V^{(0)}_{\rm eff}$, since quantum corrections to $V^{(0)}_{\rm eff}$ should be comparatively small.   

From the form of $V^{(0)}_{\rm eff}$, one might think that once $\mu > m$, the scalars should condense, breaking the $U(1)$ gauge symmetry and making the system a superconductor.  Moreover, since the masses of the electrons and selectrons are fixed to be identical due to supersymmetry, the fermions should naively start populating a Fermi surfaces at the same time that the scalars start condensing.

But there is an immediate subtlety we must deal with:  supersymmetric gauge theories typically have moduli spaces protected by supersymmetry at $\mu=0$.  In the current context, the moduli space for $m=0,\, \mu=0$ is isomorphic to $\mathbb{C}$, and is parametrized by the value of $\phi_{+} = \phi_{-}$.  For any set of vacuum expectation values for the selectrons  satisfying $\phi_{+} = \phi_{-}$, the potential energy vanishes.  But as soon as we make $\mu>m$, $V^{(0)}_{\rm eff}$ develops a runaway direction along $\phi_{+} = \phi_{-}$.  That is, the effective potential becomes unbounded from below, and the theory as defined in Eq.~\eqref{eq:sQED} does not make sense for $\mu > m$.\footnote{If both a finite $\mu$ and finite temperature $T$ are turned on, things may be different, since the finite temperature breaks supersymmetry, and should help lift the moduli space at $\mu =0$.  For an interesting recent exploration of finite-T physics in a supersymmetric gauge theory using field-theoretic techniques, see \cite{Paik:2009iz}.   }

This should not be especially surprising.  For a system comprised of weakly-interacting bosonic particles to be stable at finite chemical potential, the bosons must have sufficiently repulsive interactions.  If the interactions of the bosons were attractive, then the system would be unstable against a collapse towards arbitrarily high densities, and there would not be any equilibrium finite-density ground state.  This is precisely the issue that one faces in $\CN=\,1$ sQED, where supersymmetry demands the presence of an attractive interaction between the positive and negative selectrons $-\frac{g^2}{2} |\phi_{+}|^2 |\phi_{-}|^2$.  The arguments above imply that this issue indeed causes an instability which is unavoidable without deforming the theory in some way.

Fortunately, in $\CN=\,1$ sQED, it is possible to dodge this problem  by turning on a Fayet-Iliopulous term, which does not explicitly break the supersymmetry of the action, and has the effect of modifying the potential to
\begin{align}
\label{eq:N1ScalarPotential}
V^{(0)}_{\rm eff} = (|m|^2-\mu^2)(|\phi_{+}|^2 + |\phi_{-}|^2) + \frac{g^2}{2} (|\phi_{+}|^2 - |\phi_{-}|^2-\xi^2)^2 ,
\end{align}
where $\xi^2$ can be either positive or negative, and has mass dimension two.  At $\mu=0$, this lifts the moduli space, and indeed supersymmetry becomes spontaneously broken for $\xi>0$ so long as $m \neq 0$.  With $\xi$ turned on, we will argue that the selectrons of the theory have a stable non-trivial ground state for $\mu$ in a certain range.  Hence the naive expectation that the $U(1)$ gauge symmetry is broken at finite density is borne out, and the system is a superconductor.

One might have hoped that so long as $g \ll 1$, and the system is weakly coupled, the response of the electrons to the chemical potential should resemble that of the free limit $g=0$.  This is true in a QED plasma.  However, one should not expect it to be true in general for supersymmetric plasmas, as we now explain.   

First, it is clear from the structure of the Yukawa terms in Eq.~\eqref{eq:sQED}, which include terms of the form
\begin{align}
g\, \phi_{+}^{\dagger} \bar{\psi} P_{-} \lambda,
\end{align}
 that turning on scalar VEVs leads to mixing between the electron and gaugino fields, and this makes it difficult to guess what the fermionic fields will do in response to a chemical potential for the electrons just by looking at Eq.~\eqref{eq:sQED}.  The way to deal with this is obvious in principle, since one just has to rotate to an eigenbasis where the kinetic terms for the fermions become diagonal in the in-medium `flavors', but in practice actually doing such a rotation can be algebraically involved.  Since the coefficient of the mixing term is proportional to $g$, however, one might have hoped that when $g \ll 1$, the mixing would be small, and the response of the fermions to a chemical potential would be close to that of the $g=0$ system.

To see why this expectation is too naive, note that the coefficient of the Yukawa terms is forced to be the gauge coupling $g$ by supersymmetry.  But the coefficient controlling the strength of the self-interaction of the selectrons in Eq.~\eqref{eq:N1ScalarPotential} is $g^2$, which is also fixed by supersymmetry.   So unlike in a non-supersymmetric system, here the strengths of the Yukawa interactions and the selectron self-interactions cannot be tuned independently. In particular, given the form of the selectron potential it is obvious that a non-zero selectron VEV $\langle \phi \rangle$ must scale as
\begin{align}
\langle \phi \rangle \sim \frac{1}{g}.
\end{align}  
So since the size of the electron-gaugino mixing terms is controlled by $g \langle \phi\rangle$, we see that the fermion mixing will be essentially independent of $g$. The mixing alters the dispersion relations of the fermion fields at the quadratic level, and so we cannot assume that the response of the electrons to a chemical potential at $g=0$, which involves the formation of a Fermi surface, will necessarily persist to \emph{any} $g > 0$, no matter how small.  This observation is generic, and applies to essentially any supersymmetric gauge theory in which one turns on a chemical potential for selectrons or squarks which can also cause selectron or squark condensation.

\subsection{Summary of expectations}
\label{sec:ExpectationsSummary}
For the reasons discussed above, we expect that:
\begin{itemize}
\item The chemical potentials we will consider couple to both fermions and scalars, and so long as the theory is supersymmetric we expect the scalars to condense at the same time as the fermions begin to feel the chemical potential.  This means the $U(1)$ gauge symmetry will be broken, and the supersymmetric plasmas will be superconductors.  
\item It is essential to take into account the electric neutrality constraint.  In a related context, this was also emphasized in \cite{Huijse:2011hp}.
\item We assume that the densities are large enough that we do not have to worry about the formation of supersymmetric atoms, so that we deal with a completely ionized plasma.  This means we are focusing on the \emph{fractionalized} regime of the plasma, as opposed to the low-density atomic gas \emph{confined} regime.
\item Achieving a stable finite-density ground state may be tricky due to possible run-away directions in the scalar potential due to supersymmetry.
\item We should not expect the behavior of the fermions to be close to that of a conventional free system once there is scalar condensation, because of the structure of the Yukawa interactions and the scalar self-interactions dictated by supersymmetry.
\item If the scalars condense, the fact that the $U(1)$ electron number symmetry is shared between the scalars and the fermions means that the resulting quantum liquids will not be `compressible quantum matter' as it is defined in \cite{Huijse:2011ef}.   Moreover, the assumptions of Luttinger's theorem \cite{Luttinger:1960zz}, which ties the charge density carried by a fermionic system to the volume of the Fermi surface, will not apply to such a liquid.  So we should not expect the existence of Fermi surfaces to be automatic for finite-density supersymmetric QED.
\end{itemize}

With these observations in mind, we turn to a more detailed examination of these issues in our $\CN=\,1$ and $\CN=\,2$ sQED toy models.

\section{$\mathcal{N}=1$ sQED at finite electron number density}
\label{sec:N1QED}

\subsection{Scalar Ground State}  
\label{sec:Moduli1}
We now write down the complete $\mathcal{N} =1$ action we will consider. We include two chiral  superfields $\Phi_{+}$ and  $\Phi_{-}$ which supply the matter fields for the `electron' sector: the electron Dirac spinor field $\psi$, as well the bosonic selectrons $\phi_{+},\, \phi_{-}$.  We also include two other chiral superfields $Q_{+}$ and  $Q_{-}$, which supply the matter fields for the `ion' sector: the ion Dirac spinor field $\eta$, as well as the bosonic sion fields $q_{+},\,q_{-}$.  We consider a superpotential of the simplest possible form
\begin{equation}
\CW= m (\Phi_{+} \Phi_{-} + Q_{+} Q_{-}) ,
\end{equation}
so that the ions and the electrons have the same mass $m$.  The tree-level Kahler potential is
\begin{equation}
\CK = \Phi_+^{\dagger} e^{V} \Phi_+ + \Phi_-^{\dagger} e^{-V}\Phi_-  
+ Q_+^{\dagger} e^{V} Q_+ + Q_-^{\dagger} e^{-V}Q_- ,
\end{equation}
and $V$ is the vector superfield, which includes the photon and photino fields $A_{\mu},\, \lambda$.  We also allow a Fayet-Iliopoulos term
\begin{align}
\CL_{\xi} = -\xi^2 \int d^{4} \theta V .
\end{align}
 The Lagrangian of the version of $\CN=\,1$ SQED that we will consider is thus
\begin{align}
\CL_{\CN=\,1} = \left(\frac{1}{4g^2}\int d^2 \theta \,W^2 + \textrm{h.c.} \right)+ \int d^{4} \theta \, \CK + \left( \int d^2 \theta \, \CW + \textrm{h.c.} \right) + \CL_{\xi} ,
\end{align}
and $W_{\alpha}$ is the photon field strength chiral super field.

The matter sector has two obvious $U(1)$ symmetries, $U(1)_e$ and $U(1)_i$, which act on the component fields as shown in Table~\ref{table:SymmetriesN1}.   The diagonal $U(1)_e \times U(1)_i$ symmetry (acting as $\psi \to e^{-i\alpha} \psi,\eta \to e^{-i\alpha} \eta$, and so on) is gauged, and we will refer to the gauged charge as the `electric' charge.

\begin{table}[t]
\centering
\begin{tabular}{ c |  c  c c  c c c c }
\toprule
\multicolumn{7}{c}{Transformation Properties in $\CN=\,1$ sQED} \\
\hline
Fields: & $\psi $ & $\phi_{+}$ & $\phi_{-}$ & $\lambda$ & $\eta$ & $q_{+}$ & $q_{-}$ \\
\hline
$U(1)_e$ & $e^{-i\alpha} \psi $ & $e^{+i\alpha} \phi_{+} $ &  $e^{i\alpha} \phi_{-}$ & $\lambda$ & $ \eta $ & $q_{+}$ & $q_{-}$ \\
$U(1)_i$ & $ \psi $ & $ \phi_{+} $ & $\phi_{-}$ & $\lambda$ & $e^{-i\alpha} \eta $ & $ e^{+i\alpha} q_{+}$ & $e^{-i\alpha} q_{-}$\\
\hline
\end{tabular}
\caption{Matter field transformation properties under the $U(1)_e$ and $U(1)_i$ symmetries.}
\label{table:SymmetriesN1}
\end{table}

We want to have a net density of electron-sector fields --- electrons, selectrons, or both --- in the ground state.  To do this we turn on a chemical potential $\mu_e$ for the $U(1)_e$ symmetry, which appears in the action as the time component of a background gauge field coupling only to $U(1)_e$ charge.  At the same time, we wish to maintain charge neutrality.  To do this, we also turn on a chemical potential $\mu_i$ for the conserved charge associated with the ion $U(1)_i$ symmetry.  Then the $\mu_e$ chemical potential can be viewed as the parameter controlling the matter density of the system, while $\mu_i$ is an auxiliary parameter determined by the requirement of charge neutrality.  

It turns out that setting $\mu \equiv \mu_e = -\mu_i$ will be sufficient to maintain charge neutrality.  Heuristically, turning on $\mu > 0$ gives an equal energetic subsidy to the \emph{particles} created by the field operators $\psi,\, \phi_{-},\, \phi_{+}$ and the \emph{antiparticles} created by $\eta,\, q_{-},\, q_{+}$.  Since these two sets of particles and antiparticles have the same masses but opposite electric charges, this will create a ground state which is electrically neutral.  To see this in a more quantitative way, recall that we can read off the expression for the charge density from the part of the action which is linear in $A_0$:
\begin{align}
g A_0 Q \in \mathcal{L} ,
\end{align}
since $A_0$ is, by definition, the source for $Q$.  This yields
\begin{align}
\label{eq:QN1}
Q &= - \bar{\psi} \gamma^0 \psi +
 i[ \phi_{+}^{\dag} (\partial_0 + i \mu) \phi_{+} - ((\partial_0 + i \mu) \phi_{+})^{\dag}  \phi_{+} ] + i[ \phi_{-}^{\dag} (\partial_0 - i \mu) \phi_{-} - ((\partial_0 - i \mu) \phi_{-})^{\dag}  \phi_{-} ] \nonumber\\
 &+\bar{\eta} \gamma^{0} \eta +
 i[ q_{+}^{\dag} (\partial_0 -i \mu) q_{+} - ((\partial_0 - i \mu) q_{+})^{\dag}  q_{+} ] + i[ q_{-}^{\dag} (\partial_0 + i \mu) q_{-}  - ((\partial_0 + i \mu) q_{-})^{\dag}  q_{-} ] .
\end{align}
If $Q \neq 0$ in the ground state, the system would not be electrically neutral.  As explained above, this would not be physically sensible, since the infinite-volume limit would come with a divergent energetic cost.   More formally,  one can see that the situation when $\langle Q \rangle \neq 0$ would be problematic because then the action for $A_{\mu}$ would involve a tadpole term for $A_0$.  Once one adjusts $\mu_i$ to set $\langle Q \rangle = 0$, so that the ground state is electrically neutral, the action for $A_{\mu}$ becomes quadratic.  

We start by considering the scalar sector, and look for ground states in which the bosonic fields get time-independent vacuum expectation values, so that $\partial_0 \phi_{\pm} = \partial_0 q_{\pm} = 0$.  We use unitary gauge in our analysis, so that if any of the scalars (which are all charged under $U(1)_Q$) condense, the gauge bosons pick up a mass via the Higgs mechanism.  If two scalars condense in such a way that both $U(1)_e$ and $U(1)_i$ are broken, then one of the would-be Goldstone bosons will be eaten by the gauge field in unitary gauge, but the other will remain as a bona-fide physical gapless Goldstone mode. 

If we take $\mu_e = -\mu_i \equiv \mu$, then we get the tree-level matter sector scalar potential
\begin{equation}
V^{(0)}_{\rm eff} = \left( |m|^2 - \mu^2 \right) \left( |\phi_+|^2+ |\phi_-|^2 + |q_+|^2 + |q_-|^2 \right) 
+ \frac{g^2}{2}\left(|\phi_+|^2 - |\phi_-|^2  + |q_+|^2  - |q_-|^2  - \xi^2 \right)^2 .
\end{equation}
To develop a heuristic understanding of the scalar field ground states, it is instructive to rewrite the potential as
\begin{align}
\label{eq:SuggestivePotential1}
V^{(0)}_{\rm eff} &= \left( |m|^2 - \mu^2 - g^2 \xi^2 \right) \left(|\phi_+|^2+ |q_+|^2\right) + \left( |m|^2 - \mu^2 +g^2 \xi^2 \right)  \left(|\phi_-|^2 + |q_-|^2 \right) \nonumber \\
&+ \frac{g^2}{2}\left(|\phi_+|^2 + |q_+|^2  - |\phi_-|^2 - |q_-|^2\right)^2 + \frac{g^2}{2} \xi^4.
\end{align}
Now suppose that $\xi^2 > 0$, and consider $m^2_{\phi_{+},\, q_{+}}$ and $m^2_{\phi_{-},\, q_{-}}$ while we slowly increase $\mu$ from $0$.  (What would happen if $\xi^2< 0$ can be read off from the following discussion by exchanging $\phi_{+},q_{+}$ with $\phi_{-}, q_{-}$.)  When $m^2 - \mu^2 > g^2 \xi^2 >0$, we have $m^2_{\phi_{+}, q_{+}} > 0 $ and $m^2_{\phi_{-}, q_{-}}> 0$, so none of the scalars condense.  That is, all of the scalar VEVs are zero. This regime of the theory is not interesting for our purposes, since the scalar sector does not respond non-trivially to the chemical potential.  Moreover, given that in this regime $\mu^2 < |m|^2$ and there is no scalar condensation to leading order in $g$, the fermion sector responds to $\mu$ in the same way as a free theory would - which is to say, no spinor electrons or ions populate the vacuum either. 

Next, suppose that $-g^2\xi^2 < m^2 - \mu^2 < g^2 \xi^2$.  Then $m^2_{\phi_{+}, q_{+}} < 0$ while $m^2_{\phi_{-}, q_{-}} > 0$ , and $\phi_{+},\, q_{+}$ will develop non-trivial VEVs, and minimization of the scalar potential naively implies that they must satisfy
\begin{equation}
\label{eq:NaiveVEV1}
|\phi_+|^2 + |q_+|^2 = \frac{\mu^2 - m^2+ \xi^2 g^2}{g^2}  , \qquad |\phi_-|^2 = |q_-|^2=0 ,
\end{equation}
Plugging these VEVs back into the potential to get a feeling for what happens to $\phi_{-},\, q_{-}$, we find that  $m^2_{\phi_{-}, q_{-}}$ vanishes due to contributions from cross-terms in the potential linking $\phi_{+},\, q_{+}$ with $\phi_{-},\, q_{-}$.  This means that one should do a more careful analysis to understand the regime in which it is consistent to assume that $\phi_{+}$ and $q_{+}$ are condensed, but $\phi_{-}$ and $q_{-}$ are not.  Computing the eigenvalues $\lambda_{1}, \lambda_{2}$ of the Hessian matrix describing the fluctuations around the VEVs in Eq.~\eqref{eq:NaiveVEV1} yields
\begin{align}
\lambda_1 &= m^2 - \mu^2 - g^2 \xi^2 + g^2 ( |\phi_+|^2 + |q_+|^2 ), \\
\lambda_2 &= m^2 - \mu^2 + g^2 \xi^2 - g^2( |\phi_+|^2 + |q_+|^2 ).
\end{align}
Demanding that $\lambda_1, \lambda_2 >0$, so that our field configuration is stable, implies that we must ensure that $m^2 > \mu^2$.   Hence we learn that so long as  $0< m^2 - \mu^2 < g^2 \xi^2$, $\phi_{+}, q_{+}$ are condensed and must obey Eq.~\eqref{eq:NaiveVEV1}, but $\phi_{-}, q_{-}$ do not condense. In this regime we expect a non-trivial scalar ground state, and we do not have to worry about run-away directions in the potential.  But once $m^2 - \mu^2 < 0$, all of the scalar fields are free to develop non-zero VEVs.  Given the form of the potential, there is clearly a run-away direction in the potential along $\phi_{+} = q_{+} = \phi_{-} = q_{-}$, so the system has no stable ground state once $\mu^2>m^2$.  

Given the remarks above, we can simplify the discussion without loss of generality by assuming that $\xi^2 >0$ from here onwards.   We still have to take the constraint of charge neutrality into account.  The scalar contribution to Q is 
\begin{align}
\label{eq:ScalarCharge}
Q|_{\rm scalar} &=  2\mu_{e} |\phi_{+}|^2 -  2\mu_{e} |\phi_{-}|^2 + 2\mu_{i} |q_{+}|^2 -  2\mu_{i} |q_{-}|^2 ,
\end{align}
which becomes
\begin{align}
Q|_{\rm scalar} = 2 \mu \left( |\phi_+|^2 - |q_+|^2 \right). 
\end{align}
If we now demand that $Q|_{\rm scalar} \demand 0$, we find that
\begin{align}
\label{eq:StableVEVs1}
\boxed{|\phi_{+}|^2 = |q_{+}|^2 = \frac{\mu^2 - m^2+ \xi^2 g^2}{2 g^2}, \;\;\; \phi_{-} = q_{-} = 0.}
\end{align}
Although here we have focused on the selectrons and sions, it is clear that the symmetric way $\mu$ enters the action guarantees that if the fermionic electron and ion fields contribute to the charge density, they do so in such a way that the sum of their electric charges is separately zero.  This is the reason that we are able to demand that the scalar contribution to the electric charge vanishes separately from the one from the fermions.

\subsection{Search for a Fermi surface}
\label{sec:Fermions1}
We now examine the fermionic part of the action to see whether the fermions organize into a Fermi sphere at $\mu > 0$. Of course, in view of the discussion above, while looking for a Fermi surface, we have to always assume the condition
\begin{align}
\label{eq:MuRange}
0 <m^2 - \mu^2 < g^2 \xi^2 .
\end{align}
In particular, we emphasize that $\mu^2 < m^2$ throughout this range.  If we were to consider $\mu^2 > m^2$, the scalar sector would have no stable ground state.  On the other hand, if $\mu^2 < m^2$ but $\mu$ were to go outside the bound in Eq.~\eqref{eq:MuRange}, the scalars would have vanishing VEVs.  But then because at the same time $\mu$ would be smaller than the fermion mass, the ground state could not possibly carry any $U(1)_e$ charge.  So insisting on the condition in Eq.~\eqref{eq:MuRange} is essential to keep things interesting.

We recall that to see whether a system has a Fermi surface to leading order in perturbation theory, one can examine the dispersion relations for the fermions.  For instance, for a free Dirac fermion with Lagrangian
\begin{align}
\label{eq:FreeDirac}
\CL = \bar{\psi} \left( i\slashchar{\partial} - m + \mu \gamma^0 \right) \psi  =  \bar{\psi} M \psi ,
\end{align}
this can be done by finding the momentum-space eigenvalues $\lambda_{i}(p_0, p)$ of $M$, and then solving $\lambda_{i}(p) = 0$ for $p_0$ in terms of $p$.  This yields the dispersion relations for the fermion and anti-fermion modes determined by
\begin{align}
(p_0 - \mu)^2 = p^2 + m^2 .
\end{align}
A Fermi surface can be defined as the solution to $0= p_0 = \epsilon(p)$ for some $p = p_F >0$. For a free fermion, we of course obtain $p^2_F = \mu^2 - m^2$.   Our task in this section is to carry out this simple procedure for the somewhat baroque fermion sector of sQED.

In four-component spinor notation, the fermion part of the $\CN = 1$ sQED Lagrangian is
\begin{align}
\label{eq:Fermions1}
\mathcal{L}_{\CN=\,1} |_{\rm fermion} &= 
\frac{1}{2}\bar{\lambda} i\slashchar{\partial} \lambda + \bar{\psi}\left(i\slashchar{D}^{-} - m\right) \psi  +  \bar{\eta}\left(i\slashchar{D}^{-} - m\right) \eta  \nonumber \\
&+ \sqrt{2} i g \left( \phi_{+}^{\dagger}\bar{\psi} P_{-} \lambda - \phi_{-}^{\dagger} \bar{\lambda} P_{-} \psi - \phi \bar{\lambda} P_{+} \psi   + \phi_{-} \bar{\psi} P_{+} \lambda    \right)\\
&+ \sqrt{2} i g \left( q_{+}^{\dagger}\bar{\eta} P_{-} \lambda - q_{-}^{\dagger} \bar{\lambda} P_{-} \eta - q \bar{\lambda} P_{+} \eta   + q \bar{\eta} P_{+} \lambda    \right) ,  \nonumber
\end{align}
where
\begin{align}
D^{-}_{\mu} \psi &= \partial_{\mu} - i \mu \delta_{\mu, 0} - i g A_{\mu} \\
D^{-}_{\mu} \eta &= \partial_{\mu} + i \mu \delta_{\mu, 0} - i g A_{\mu}
\end{align}
In view of our discussion in Section~\ref{sec:Challenges} and the response of the scalar sector to the chemical potential, once the scalar fields develop non-trivial VEVs in Eq.~\eqref{eq:StableVEVs1} all of the fermionic fields mix with each other, with the mixing between electron and ion fields mediated by the photino.   Moreover, if for simplicity we scale $\xi$ as $\xi \sim 1/g$, the mixing is $g$-independent.  It is thus difficult to understand the response of the fermions to the chemical potential through a visual examination of Eq.~\eqref{eq:Fermions1}, in contrast to the free case in Eq.~\eqref{eq:FreeDirac}.

To look for a Fermi surface, we want to compute the dispersion relations of the fermionic eigenmodes described by Eq.~\eqref{eq:Fermions1}.    This is easier if we switch to two-component spinor notation, where
\begin{align}
\psi =\left(\begin{array}{c}
\psi_{L\alpha} \\
\psi_R^{\dagger\dot{\alpha}} \\
\end{array}\right), \;\;\; \eta =\left(\begin{array}{c}
\eta_{L\alpha} \\
\eta_R^{\dagger\dot{\alpha}} \\
\end{array}\right) ,
\end{align}
the Majorana photino is written as 
\begin{align}
\lambda =  \left(\begin{array}{c}
\lambda_\alpha \\
\lambda^{\dagger\dot{\alpha}} \\
\end{array}\right) ,
\end{align}
and we introduce the standard matrices $\sigma^{\mu}_{\alpha \dot{\alpha}} = \left(I_2, \,  \vec{\sigma} \right)$ and $\bar{\sigma}^{\mu  \dot{\alpha}\alpha} = \left(I_2, \, -\vec{\sigma} \right)$. 

The fact that the VEVs are given by Eq.~\eqref{eq:StableVEVs1} means that one can write the quadratic fermion action in terms of a $5\times5$ matrix, without the need to introduce Nambu-Gorkov-type spinors. Defining 
\begin{align}
\Psi =\begin{pmatrix}\psi_{L \alpha} \cr \psi_{R}^{\dagger\dot{\alpha}}\cr \lambda_{\alpha} \cr \eta_{L\alpha} \cr \eta_{R}^{\dagger \dot{\alpha}}  \end{pmatrix}
\end{align}
we can now write
\begin{align}
\mathcal{L}_{\CN=\,1} |_{\rm fermion} &= \Psi^{\dag} \cdot M_{\CN =1} \cdot \Psi,
\end{align}
where
 \begin{align}
M_{\CN =1} = \begin{pmatrix}
 i \bar{\sigma}^{\mu}\left(\partial_{\mu}-i\mu \delta_{\mu 0}\right)& -m I_2 &0 &0 &0 \cr 
-m I_2 &i \sigma^{\mu}\left(\partial_{\mu}-i\mu \delta_{\mu 0}\right) & i g \sqrt{2} \phi_+^{\dagger} &0 &0 \cr 
0 & -i g \sqrt{2}  \phi_+ & i \bar{\sigma}^{\mu}\partial_{\mu} &0 &- i g \sqrt{2} q_+ \cr
 0 &0 &0 & i \bar{\sigma}^{\mu}\left(\partial_{\mu}+i\mu \delta_{\mu 0}\right)& -m I_2 \cr 
 0 & 0& i g \sqrt{2} q_+^{\dagger} &  -m I_2 &i \sigma^{\mu}\left(\partial_{\mu}+i\mu \delta_{\mu 0}\right)   
  \end{pmatrix} .
\label{eq:M1}
\end{align}
After going to momentum space we can compute the determinant of $M_{\CN =1}$. Lorentz invariance is broken by $\mu$, but rotational invariance is unbroken and hence $\det(M_{\CN =1})$ must depend on $p_0$ and $p= \sqrt{p_1^2+p_2^2 + p_3^2}$.  The dispersion relations may be found by solving $\det(M_{\CN =1}) = 0$ for $p_0$ as a function of $p$, but they are ugly enough that we do not show their general form, which seems unilluminating.  Fortunately, once we set $p_0 = 0$, as is needed in the search for the Fermi surface, things become prettier and we find
\begin{align}
\det(M_{\CN =1})|_{p_0 = 0} &= \left(-\mu ^2+m^2+p^2\right)^2 \left(2 g^2 (\mu +p) \left(\left| \phi_{+} \right| ^2-\left|q_{+}\right|^2\right)+p \left(-\mu ^2+m^2+p^2\right)\right) \\
&\times \left(p \left(\mu^2-m^2-p^2\right)+ 2 g^2 (p-\mu ) \left(\left|q_{+}\right| ^2-\left|\phi_{+} \right|^2 \right)\right) \\
& = - p^2 \left(-\mu ^2+m^2+p^2\right)^4 .
\end{align}
Note that once the smoke has cleared, the contribution of the selectron and sion VEVs to $\det(M_{\CN =1})|_{p0 = 0}$ cancels thanks to charge neutrality.  Amusingly, what is left has the form which we would have obtained by dropping the Yukawa terms in the first place!  We emphasize that this dramatic simplification happens only at $p_0 = 0$.

 Looking for values of $p = p_F > 0$ which make $\det(M_{\CN =1})|_{p_0 = 0}$ vanish, at first glance $p = \sqrt{\mu^2 - m^2}$ may seem to do the job.  But as we have seen, the scalar sector is under control only for $\mu < m$, and indeed we have assumed the condition in  Eq.~\eqref{eq:MuRange} at the start of the fermion analysis.  So  $p = \sqrt{\mu^2 - m^2}$ is \emph{not} a legitimate solution of $\det(M_{\CN =1})|_{p_0 = 0} = 0$.  But there are no other solutions to $\det(M_{\CN =1})|_{p_0 = 0} = 0$.
 
Thus we conclude that within the domain of validity of our analysis, there is no $p = p_F >0$ for which $\det(M_{\CN =1})|_{p_0 = 0}$ vanishes, and hence there is no Fermi surface in finite-density $\mathcal{N}=1$ sQED at weak coupling.  Note also that changing the strength of the Yukawa couplings (which would break supersymmetry) would not change this result due to the structure of the determinant above.

\subsection{Non-supersymmetric cousin of $\CN=\,1$ sQED}
\label{sec:N1HardBreaking}
Before proceeding to $\CN=\,2$ sQED, it is instructive to discuss what would have happened if we had not insisted on charge neutrality, for instance by working with only the electron-sector fields. The point of considering this example is to emphasize that $U(1)$ breaking does \emph{not} necessarily lead to the obliteration of Fermi surfaces. One way to make this reasonable would be to modify the Lagrangian by erasing the gauge field while leaving everything else untouched. Then the Lagrangian would be
\begin{align}
\mathcal{L}_{\textrm{no ions}} = \mathcal{L}_{\CN=\,1} |_{\rm fermion}  + \left|D^{-}_{\mu} \phi_{-}\right|^{2}+\left|D^{+}_{\mu} \phi_{+}\right|^{2}  - \left|m \phi_{-}\right|^{2}-\left|m\phi_{+}\right|^{2} - \frac{g^{2}}{2} \left( \left|\phi_{+}\right|^{2} - \left|\phi_{-}\right|^{2} \right)^{2}
\end{align}
with the ion fields deleted from $\mathcal{L}_{\CN=\,1} |_{\rm fermion}$.  Deleting the gauge fields breaks supersymmetry.

\begin{figure}[htbp]
\centering
\includegraphics[width=0.48\textwidth]{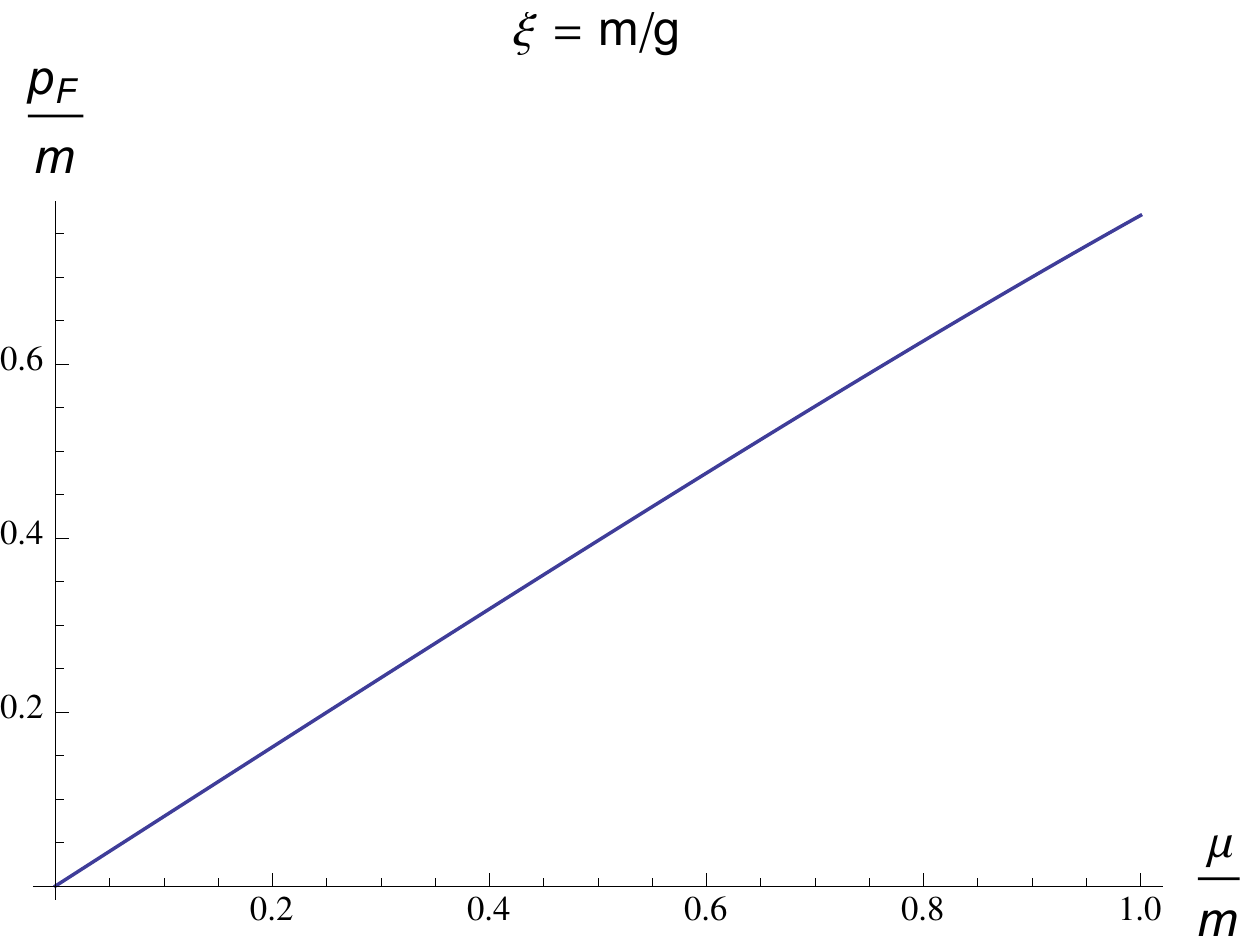}
\caption{A solution for the Fermi momentum $p_F$ which would have been obtained if we had ignored the constraint of charge neutrality, and worked with a system including electrons only.  For simplicity we set $\xi = m/g$.  In this case the infinite volume does not make sense, unless one modifies the theory by removing the gauge fields but keeping all else fixed. }
\label{fig:PfNoIons}
\end{figure}

Going through the same analysis as above, we now obtain
\begin{align}
\det(M_{\textrm{no ions}})_{p_0 = 0} &= 4 g^4 \left| \phi_{+} \right| ^4 \left(\mu ^2-p^2\right)-p^2 \left(-\mu ^2+m^2+p^2\right) \left(4 g^2 |\phi_{+}|^2-\mu ^2+m^2+p^2\right)
\end{align}
with $g^2 |\phi_{+}|^2 = m^2 -\mu^2 + g^2 \xi^2$.  Solving $\det(M_{\textrm{no ions}})|_{p_0 = 0} = 0$, we obtain a solution for the Fermi momentum:
\begin{align}
p_F &= \frac{\left[c+27 \mu  \left(g^2 \xi ^2-\mu ^2+m^2\right)\right]^{2/3}-6 g^2 \xi ^2+9 (\mu^2- m^2)}{3 \left[c+27 \mu  \left(g^2 \xi ^2-\mu ^2+m^2\right)\right]^{1/3}} ,
\label{eq:PfNoIons}
\end{align} 
where 
\begin{align}
c &\equiv  \sqrt{\left(6 g^2 \xi ^2-9 \mu ^2+9 m^2\right)^3+729 \mu ^2 \left(g^2 \xi ^2-\mu ^2+m^2\right)^2} .
\end{align}
Since these expressions are rather ugly, we plot Eq.~\eqref{eq:PfNoIons} is shown in Fig.~\ref{fig:PfNoIons}.   The plot shows this non-supersymmetric system \emph{does} have a Fermi surface, in contrast to the supersymmetric system we considered above.  Note that the Yukawa terms are essential to this result, since here we are still considering $\mu<m$, so that without the mixing terms the fermions would be free to leading order, and would not develop a Fermi surface until $\mu>m$.

However, as we have seen, when electric neutrality is taken into account, as it \emph{must} be in $\CN=\,1$ sQED, the story is very different.

\section{Softly broken $\mathcal{N}=2$ sQED at finite electron number density}
\label{sec:N2QED}

\subsection{Scalar Ground State}  
\label{sec:Moduli2}
We start by attempting to work with the most obvious $\CN=\,2$ generalization of our $\CN=\,1$ toy model.  As the field content of our $\CN=\,2$ sQED model, we will use  essentially the same chiral `electron' and `ion' super fields as the $\CN=\,1$ model, with the following changes.  First, we must add an extra `adjoint'  $\CN=\,1$ chiral multiplet $\Lambda$ which contains an extra Majorana photino $\chi$ and a scalar $a$, which combines with the $\CN=\,1$ vector multiplet to form the $\CN=\,2$ vector hypermultiplet.  Second, the scalar fields from the $\CN=\,1$ chiral multiplets, $\phi_{+}$ and $\phi_{-}^{\dagger}$, combine to form a single $\CN=\,2$ matter hypermultiplet.\footnote{The hypermultiplet contains the \emph{conjugate} of $\phi_-$ since the gauge generators commute with the supersymmetry generators and hence all fields within a multiplet have the same gauge charges.} The same goes for the sion fields.  Finally, to be consistent with $\CN=\,2$ supersymmetry, the superpotential must be modified to (in $\CN=\,1$ language)
\begin{align}
\CW = m\left( \Phi_+ \Phi_- + Q_+ Q_-  \right)+ \sqrt{2} \Lambda \left(  \Phi_+ \Phi_- + Q_+ Q_-  \right),
\end{align}
where $\Phi_+,\, \Phi_-$ are the electron-multiplet superfields and $Q_+,\, Q_-$ are the ion-sector superfields. The tree-level Kahler potential is the same as before with the obvious changes to account for the discussion above.  We continue to include the FI term in the theory.  This $\mathcal{N}=2$ gauge theory has the scalar potential
\begin{align}
V^{(0)}_{\rm eff} &= \left| \sqrt{2} g a +  m \right|^2   \left(\left| \phi_+ \right|^2 + \left| \phi_- \right|^2 +\left| q_{+} \right|^2 + \left| q_{-} \right|^2  \right)  
+  2 g^2 \left( \phi_+ \phi_- + q_+ q_- \right) \left( \phi_{+}^{\dagger} \phi_{-}^{\dagger}  + q_+^{\dagger}  q_- ^{\dagger} \right) \nonumber \\
&+ \frac{g^2}{2} \left( \left| \phi_+ \right|^2 - \left| \phi_- \right|^2 +\left| q_+ \right|^2 -\left| q_- \right|^2 - \xi  \right)^2 
-\mu^2  \left( \left| \phi_+ \right|^2 + \left| \phi_- \right|^2 +\left| q_+ \right|^2 +\left| q_- \right|^2  \right) ,
\label{eq:N2Potential}
\end{align}
and has the same $U(1)_e \times U(1)_i$ symmetry as the $\CN=\,1$ theory, but also has an $SU(2)_R$ non-anomalous R-symmetry.  We explore the response of $\CN=\,2$ sQED to an R-charge chemical potential in Section~\ref{sec:Rcharge}, and focus on the $U(1)_e \times U(1)_i$ symmetries here.  The transformation properties of the matter fields are given in Table~\ref{table:SymmetriesN2}.  Recalling the comments about the way $\phi_{+}$ and $\phi_{-}^{\dagger}$ enter the $\CN=\,2$ theory above, and noting  that the fields $a$ and $\xi$ do not contribute to the electric charge density, we find that the gauged (electric) charge density is unchanged from  Eq.~\eqref{eq:QN1}, 
\begin{align}
Q = &- \bar{\psi} \gamma^0 \psi +
 i[ \phi_{+}^{\dag} (\partial_0 + i \mu) \phi_{+} - ((\partial_0 + i\mu) \phi_{+})^{\dag}  \phi_{+} ] + i[  \phi_{-}^{\dag} (\partial_0 - i \mu) \phi_{-} - ((\partial_0 - i\mu) \phi_{-})^{\dag}  \phi_{-}  ] \nonumber\\
 &+\bar{\eta} \gamma^{0} \eta +
 i[ q_{+}^{\dag} (\partial_0 - i \mu) q_{+}  - ((\partial_0 - i \mu) q_{+})^{\dag}  q_{+} ] + i[q_{-}^{\dag} (\partial_0 + i \mu) q_{-}  - ((\partial_0 + i \mu) q_{-})^{\dag}  q_{-}] .
 \label{eq:QN2}
\end{align}

\begin{table}[t]
\centering
\begin{tabular}{ c |  c  c  c c c c c c c}
\toprule
\multicolumn{10}{c}{Transformation Properties in $\CN=\,2$ sQED} \\
\hline
Fields: & $\psi $ & $\phi_{+}$ & $\phi_{-}$ & $\eta$ & $q_{+}$ & $q_{-}$ & $\lambda$ & $\chi$ & $a$ \\
\hline
$U(1)_e$ & $e^{-i\alpha} \psi $ & $e^{+i\alpha} \phi_{+} $ & $e^{-i\alpha} \phi_{-}$ & $ \eta $ & $q_{+}$ & $q_{-}$ & $\lambda$ & $\chi$ & $a$ \\
$U(1)_i$ & $ \psi $ & $ \phi_{+} $ & $\phi_{-}$ & $e^{-i\alpha} \eta $ & $ e^{+i\alpha} q_{+}$ & $e^{-i\alpha} q_{-}$ & $\lambda$ & $\chi$ & $a$\\
\hline
\end{tabular}
\caption{Matter field transformation properties under the $U(1)_e$ and $U(1)_i$ symmetries.}
\label{table:SymmetriesN2}
\end{table}

Unfortunately, it turns out that once $\mu$ is turned on the scalars do not have a stable ground state, since there are run-away directions in the scalar potential.  The quickest way to see this is to observe that minimizing $V_{\rm eff}$ for $a$ implies that
 $a$ picks up a VEV 
\begin{align}
\label{eq:aVEV}
\langle a\rangle = -\frac{m}{\sqrt{2} g} . 
\end{align}
Heuristically, apart from the surviving group of terms in the first line of Eq.~\eqref{eq:N2Potential}, the potential for $\phi_{1,2},\, q_{1,2}$ is the same as the massless limit of the potential in the $\CN=\,1$ case, for which there would be no stable solutions once $\mu>0$, even when a FI term is present. The new terms demanded by $\CN=\,2$ do not save the day if there is more than one flavor hypermultiplet.   

We have not figured out a way to prevent the emergence of run-away directions in the scalar potential in two-flavor $\CN=\,2$ sQED, but it is possible to get some insight into what the supersymmetric interactions do to Fermi surfaces by modifying the theory above in two simple ways:

\vbox{%
	\begin{enumerate}[label=\Alph*:]
		\item{Work with $\CN=\,2$ sQED with only one flavor.  This means giving up on electric neutrality, and requires a hard breaking of supersymmetry to be sensible in the infinite-volume limit, much as in Section~\ref{sec:N1HardBreaking}.  We defer a discussion of this case in Section~\ref{sec:N2HardBreaking}. }
		\item{Keep the ion fields, but add some soft SUSY-breaking terms.}  
	\end{enumerate}
}

Given the title of this section, we proceed with option B, and work with a theory defined by
\begin{align}
\CL = \CL_{\CN=\,2} + m_s^2\left(|\phi_{+}|^2+|\phi_{-}|^2+|q_{+}|^2+|q_{-}|^2 \right) ,
\end{align}  
where $\CL_{\CN=\,2}$ is the Lagrangian of $\CN=\,2$ sQED with electron and ion superfields we presented above, and  $m_s$ is the soft SUSY-breaking mass. 

Minimizing the softly-broken scalar potential with respect to $a$, we again get $\langle a\rangle = -\frac{m}{\sqrt{2} g}$.  The condition for the remaining scalars to have a stable condensate is 
\begin{align}
\label{eq:SoftBreakingCondition}
m_s^2 - g^2 \xi^2 < \mu^2 < m_s^2 , 
\end{align}
where $m_s$ is the soft mass we introduced above. If the lower bound is violated none of the scalars condense, while if the upper bound is violated there is a runaway direction. If Eq.~\eqref{eq:SoftBreakingCondition} is satisfied, the scalar VEVs must obey the relations
\begin{equation}
\left|\phi_+ \right|^2 + \left|q_+ \right|^2 = \frac{\mu^2 - (m_s^2-g^2 \xi^2)}{g^2}, \; \; |\phi_{-}| = |q_{-}| = 0 .
\end{equation}
Taking into account the electric neutrality constraint means that the scalar VEVs become
\begin{equation}
\boxed{\left|\phi_+ \right|^2 = \left|q_+ \right|^2 = \frac{\mu^2 - (m_s^2-g^2 \xi^2)}{2g^2}, \; \; |\phi_{-}| = |q_{-}| = 0 .}
\label{eq:ScalarVEVsN2Broken}
\end{equation}

\subsection{Search for a Fermi surface}
\label{sec:Fermions2}
The fermionic terms in the Lagrangian are the same as in Eq.~\eqref{eq:Fermions1} together with the additional terms
\begin{align}
\label{eq:Fermions2}
\mathcal{L}_{\CN=\,2} |_{\rm fermions} &= \mathcal{L}_{\CN=\,1} |_{\rm fermions}+ \frac{1}{2}\bar{\chi} i\slashchar{\partial} \chi 
- \sqrt{2} g \left( a \bar{\psi} P_- \psi + a^{\dagger} \bar{\psi}P_+\psi +a \bar{\eta} P_- \eta + a^{\dagger} \bar{\eta}P_+\eta  \right) \nonumber  \\
&- \sqrt{2}  g \left( \phi_{-}\bar{\psi} P_{-} \chi + \phi_{+} \bar{\chi} P_{-} \psi + \phi_{-}^{\dagger} \bar{\chi} P_+ \psi + \phi_+^{\dagger} \bar{\psi} P_+\chi     \right) \\
&- \sqrt{2}  g \left( q_{-}\bar{\eta} P_{-} \chi + q_{+} \bar{\chi} P_{-} \eta + q_{-}^{\dagger} \bar{\chi} P_+ \eta + q_+^{\dagger} \bar{\eta} P_+\chi     \right) . \nonumber
\end{align}
Again, once the scalars pick up VEVs, all of the fermionic fields mix with each other, and seeing the effect of the chemical potential requires diagonalizing the kinetic operator.  To look for a Fermi surface, paralleling the approach of Section~\ref{sec:Fermions1}, we introduce a single column vector collecting all of our two-component spinors
\begin{align}
\Psi =\begin{pmatrix}\psi_{L \alpha} \cr \psi_{R}^{\dagger\dot{\alpha}}\cr \lambda_{\alpha} \cr \chi^{\dagger \dot{\alpha}} \cr \eta_{L\alpha} \cr \eta_R^{\dagger \dot{\alpha}} \end{pmatrix} .
\end{align}
This allows us to rewrite Eq.~\eqref{eq:Fermions2} as
\begin{align}
\mathcal{L}_{\CN=\,2} |_{\rm fermion} &= \Psi^{\dag} \cdot M_{\CN =\,2} \cdot \Psi, 
\end{align}
where
\begin{align}
M_{\CN =\,2} =  
\begin{pmatrix}
	i \bar{\sigma}^{\mu}\left(\partial_{\mu}-i\mu \delta_{\mu 0}\right)& 0 &0 &- g\sqrt{2} \phi_{+}^{\dagger} & 0&0 \cr 
	0 &i \sigma^{\mu}\left(\partial_{\mu}-i\mu \delta_{\mu 0}\right) & ig\sqrt{2}  \phi_+^{\dagger} &0 &0 &0  \cr 
	0 & -i g\sqrt{2}  \phi_+ & i \bar{\sigma}^{\mu}\partial_{\mu} & 0 &0 &-ig\sqrt{2} q_+ \cr 
	-g\sqrt{2}  \phi_+ &0 &0 & i\sigma^{\mu}\partial_{\mu} &-g\sqrt{2} q_+ &0 \cr
	0 &0 &0 &-g\sqrt{2} q_+^{\dagger} & i \bar{\sigma}^{\mu}\left(\partial_{\mu}+i\mu \delta_{\mu 0}\right) &0  \cr
	0 &0 &i g\sqrt{2} q_+^{\dagger}  &0 &0 &i \sigma^{\mu}\left(\partial_{\mu}+i\mu \delta_{\mu 0}\right) 
\end{pmatrix}  .
\end{align}

Going to momentum space, calculating $\det(M_{\CN=\,2})$ and setting $p_0 = 0$, we obtain
\begin{align}
\det(M_{\CN=\,2})|_{p_0 = 0} &= p^2 \left(4 g^2 |\phi_{+}|^2-\mu ^2+p^2\right)^2 \nonumber \\
&\times \left(2 g^2 |\phi_{+}|^2 (\mu +p)+(p-\mu ) \left(2 g^2 |q_{+}|^2+p (\mu +p)\right)\right) \nonumber \\
&\times \left((p-\mu ) \left(2 g^2 |\phi_{+}|^2+p (\mu +p)\right)+2 g^2 |q_{+}|^2 (\mu +p)\right) ,
\end{align}
where we have used the charge neutrality relation between the scalar VEVs.  If the scalars condense, we can plug in Eq.~\eqref{eq:ScalarVEVsN2Broken} to get
\begin{align}
\det(M_{\CN=\,2})|_{p_0 = 0}  &= p^4 \left(4 g^2|\phi_{+}|^2-\mu ^2+p^2\right)^4 \nonumber \\
&=p^4 \left(2 [\mu^2 -(m_s^2- g^2 \xi^2)] -\mu ^2+p^2\right)^4 \nonumber \\
&=p^4 \left(\mu^2 - 2 m_s^2+  2 g^2 \xi^2+p^2\right)^4   .
\end{align}
Looking for a value of $p \neq 0$ which would make this vanish, we find that $p_F$ would have to satisfy the relation 
\begin{equation}
p^2_f =  2m_s^2 - \mu^2 - 2g^2 \xi^2 \overset{!}{>} 0.
\end{equation}
This relation will be satisfied if
\begin{equation}
\mu^2< 2( m_s^2 - g^2 \xi^2) .
\end{equation}

\begin{figure}[htbp]
\centering
\includegraphics[width=0.48\textwidth]{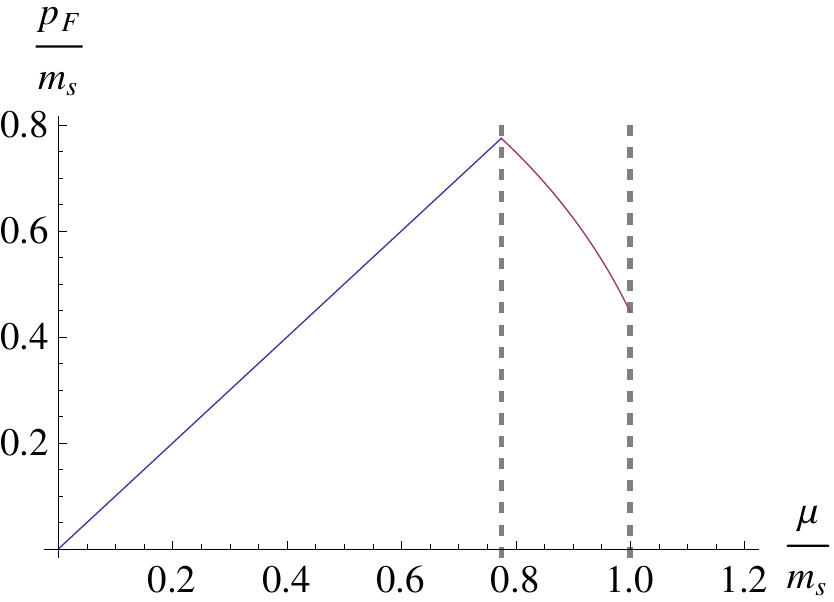}
\includegraphics[width=0.48\textwidth]{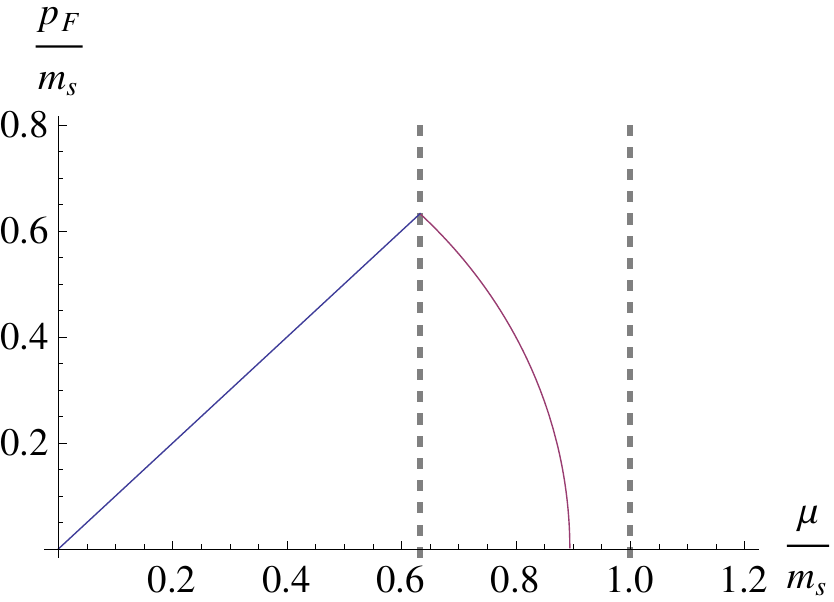}
\caption{{\bf Left:} Fermi momenta as a function of $\mu$ with $\frac{g^2 \xi^2}{m_s^2}  = 0.4$. {\bf Right:} Fermi momenta as a function of $\mu$ with $\frac{g^2 \xi^2}{m_s^2} = 0.6$. The area between the dashed lines is the region where the scalars are condensed and stable. Values of $\mu$ to the right of this region make the scalars unstable, while to the left, the scalars are not condensed.   Note that past $g^2 \xi^2 > m_s^2$ , where the scalars are always condensed, there is no Fermi surface.}
\label{fig:Pfsoft}
\end{figure}

We are now in a position to classify all the things that can happen to the fermions in this theory.  To begin with, if
\begin{equation}
 \mu^2 < m_s^2 - g^2 \xi^2 ,
\end{equation}
then the charged scalars do not condense.  The fermion sector consists of massless gauginos and massless matter fermions, which to leading order are free. Since the matter fermions feel the chemical potential, there is a fermi surface at $p_F = \mu$.  Since the charged scalars are not condensed, the system is not a superconductor (before considering fermion pairing effects), and it is natural to speculate that the physics in this regime resembles that of conventional QED plasmas.

Next, if
\begin{align}
m_s^2 - g^2 \xi^2 < \mu^2 < 2m_s^2 - 2g^2 \xi^2 \;\; \textrm{and} \;\; \mu^2 <  m_s^2 ,
 \end{align}
the theory is stable, the charged scalars are condensed, so that the quantum liquid is a superconductor, and there is a Fermi surface.

If
\begin{align}
2m_s^2 - 2g^2 \xi^2 < \mu^2< m_s^2  ,
 \end{align}
the scalar sector is stable, with the charged scalars condensed and hence a broken $U(1)_Q$, so that the system is a superconductor.   But now there is no Fermi surface.

Finally if 
\begin{align}
m_s^2 < \mu^2  ,
 \end{align}
the scalar sector becomes unstable, and there does not appear to be a sensible finite-density ground state.

To help visualize the behavior of the Fermi surfaces in this theory as a function the parameters, see Fig~\ref{fig:Pfsoft}.  As seen in the plots, as the scalar condensates get larger, the Fermi momentum decreases.  Naively, one could interpret Fig.~\ref{fig:Pfsoft} as implying that more and more of the charge in the system leaks from the fermions into the scalars as $\mu$ is increased enough to make the scalar condensate start growing.  But see Section~\ref{sec:RChargeFermions} for a result which suggests that this is not necessarily the case.

\subsection{Non-supersymmetric cousin of $\CN=\,2$ sQED}
\label{sec:N2HardBreaking}
We now briefly return to Option A from Section~\ref{sec:Moduli2}, where we start with $\CN=\,2$ sQED with one matter hypermultiplet, and delete the gauge fields just as in Section~\ref{sec:N1HardBreaking} to avoid problems with electric neutrality. This is a hard breaking of supersymmetry.

The scalar potential is now
\begin{align}
V^{(0)}_{\rm eff} &= \left| \sqrt{2} g a +  m \right|^2   \left(\left| \phi_+ \right|^2 + \left| \phi_- \right|^2  \right)  
+  2 g^2|\phi_+|^2 |\phi_-|^2  \nonumber \\
&+ \frac{g^2}{2} \left( \left| \phi_+ \right|^2 - \left| \phi_- \right|^2  - \xi  \right)^2 
-\mu^2  \left( \left| \phi_+ \right|^2 + \left| \phi_- \right|^2 \right) .
\label{eq:N2PotentialNN}
\end{align}
The VEV of $a$ is still given by Eq.~\eqref{eq:aVEV}, but now there is a stable minimum for the other scalar fields as well, as can be seen by rewriting the potential in the manner of Eq.~\eqref{eq:SuggestivePotential1}.   If $\xi^2 > 0$, minimizing $V^{(0)}_{\rm eff}$ leads to 
\begin{equation}
\left| \phi_+ \right|^2 = \frac{\mu^2 + g^2\xi^2}{g^2} , \qquad \left| \phi_- \right|^2=0 ,
\end{equation}
while if $\xi^2 < 0$, we get 
\begin{equation}
\left| \phi_- \right|^2  = \frac{\mu^2-g^2 \xi^2}{g^2} , \qquad \left| \phi_+ \right|^2 =0 .
\end{equation}
(At $\xi = 0$, both scalar fields can condense, but for simplicity we do not consider this case further.)   As we have been saying, in this case there is no way to solve the charge neutrality constraint within the scalar sector.  If it were possible to  adjust the chemical potential which couples to the electrons  \emph{independently} from the one which couples to the selectrons, one could imagine that this electron chemical potential could be dialed in such a way that the electrons would carry a charge density which precisely compensates that of the scalars.  But the structure of our supersymmetric theory does not allow us to introduce such an independent chemical potential for the electrons, because the Yukawa interactions do not respect the $U(1)$ electron-number symmetry of the free action.  

Hence, the solutions obtained in this section cannot yield an electrically-neutral  background.  Of course, since we have deleted the gauge fields from the theory with malice aforethought, this is not a problem. 

We now start the search for a Fermi surface for this non-supersymmetric theory.   Again, the diagonalization of the fermion sector after scalar condensation is much easier if we switch to two-component notation.  So long as $\xi^2 \neq 0$, $\mathcal{L}_{\CN=\,2} |_{\rm fermions}$ can be written in a matrix notation without introducing Nambu-Gorkov spinors, but at $\xi = 0$ we expect all of the scalar fields to develop non-trivial VEVs, making the analysis more involved.  To keep things as simple as possible, we  only discuss the $\xi^2 \neq 0$ case in this paper.  Moreover, as our previous discussion makes clear, to understand what happens for $\xi^2 \neq 0$ we can focus on $\xi^2 > 0$.

Paralleling the approach of Section~\ref{sec:Fermions1}, we introduce a single column vector collecting all of the two-component spinors
\begin{align}
\Psi^{(1)} = \begin{pmatrix}\psi_{L \alpha} \cr \psi_{R}^{\dagger\dot{\alpha}}\cr \lambda_{\alpha} \cr \chi^{\dagger \dot{\alpha}} \end{pmatrix} .
\end{align}
We rewrite Eq.~\eqref{eq:Fermions2} as
\begin{align}
\mathcal{L}_{\CN=\,2} |_{\rm fermion} &= [\Psi^{(1)}]^{\dag} \cdot M^{(1)}_{\CN =2} \cdot \Psi^{(1)} ,
\end{align}
with
\begin{align}
M^{(1)}_{\CN=\,2} = \begin{pmatrix} i \bar{\sigma}^{\mu}\left(\partial_{\mu}-i\mu \delta_{\mu 0}\right)& 0 &0 & -g\sqrt{2} \phi_{+}^{\dagger} \cr 0 &i \sigma^{\mu}\left(\partial_{\mu}-i\mu \delta_{\mu 0}\right) & ig\sqrt{2}  \phi_+^{\dagger} &0  \cr 0 & -i g\sqrt{2}  \phi_+ & i \bar{\sigma}^{\mu}\partial_{\mu} & 0 \cr -g\sqrt{2}  \phi_+ &0 &0 & i\sigma^{\mu}\partial_{\mu} \end{pmatrix} .
\end{align}
Computing the determinant of $M^{(1)}_{\CN=\,2}$ in frequency-momentum space, we find that the dispersion relations for the fermions are 
\begin{align}
p_0 = \frac{1}{2} \left(-\mu \pm \sqrt{8 g^2 \phi_+^2+4 p^2 \pm 4 p \mu +\mu ^2} \right) .
\end{align}
But one can now check that there is no value of $p^2=p^2_F > 0$ such that there is a solution to the equation above for $p_0 = 0$.  Thus there is no Fermi surface if we work with the non-electrically-neutral state in the $\CN=\,2$ theory with only one flavor hypermultiplet, or in the healthy but non-supersymmetric theory with the gauge fields removed.  Note the contrast of this result with what we saw in Section~\ref{sec:N1HardBreaking}, where the analogous theory had a Fermi  surface.

\section{$\CN=\,2$ sQED with a finite $R$-charge density}
\label{sec:Rcharge}
In this section, we will consider $\CN=\,2$ sQED with \emph{one} matter hypermultiplet. As we mentioned in the previous section, $\CN=\,1$ sQED has a $U(2) = U(1)_\CR \times SU(2)_R $ R-symmetry group. The $U(1)_\CR$ subgroup is anomalous, whereas the $SU(2)_R$ remains anomaly free. We focus on the anomaly-free symmetry.  
The $SU(2)_R$ symmetry acts by matrix multiplication on the Weyl doublet $(\lambda_{\alpha}, \chi_{\alpha})$ from the vector hypermultiplet and the charged scalars $(\phi_+, \phi^{\dag}_-)$ from the matter hypermultiplet. The remaining fields in the theory are $SU(2)_R$ singlets.

We can describe a system with a net R-charge by introducing a set of chemical potentials $\mu_{n}$ for the R-symmetry charges. Any conserved charges $Q_n$ that one wishes to introduce into the grand canonical partition function change the  Hamiltonian by a shift 
\begin{align}
\CH \to \CH - \sum_n \mu_n Q_n .
\end{align}
However, the $Q_n$ charges must commute with each other in order to be simultaneously observable.  This means that $Q_n$ can only belong to the maximally commuting (Cartan) sub-algebra of the non-Abelian algebra of the charge operators. In our case this must pick a single $U(1)_R \subset SU(2)_R$ to which we associate the chemical potential $\mu_R$. Furthermore, since this is a global un-gauged symmetry we do not have to worry about making the system neutral with respect to $U(1)_R$.  Of course, we still have to make sure we maintain electric neutrality!

Define the $SU(2)_R$ doublet fields
\begin{align}
\Phi &\equiv \begin{pmatrix}\phi_{+} \cr \phi^{\dag}_{-}  \end{pmatrix}, \;\;\;    \Psi_{\alpha} \equiv \begin{pmatrix}\lambda_{\alpha} \cr \chi_{\alpha} \end{pmatrix} .
\end{align}
Our anomaly-free $U(1)_R$ subgroup acts on these fields as 
\begin{align}
\Phi \to e^{i\alpha \tau_3} \Phi, \;\;\;   \Psi_{\alpha} \to  e^{i\alpha \tau_3} \Psi_{\alpha} ,
\end{align}
where $\tau_3 = \sigma_3$, the diagonal Pauli matrix.

Hence, the $\mu_R$ chemical potential enters the Lagrangian in the following way
\begin{align}
\CL = (\Psi_{\alpha})^{\dag} \sigma^{\mu} D_{\mu} \Psi_{\alpha} + |D_{\mu} \Phi|^2 + \ldots ,
\end{align}
where we define\footnote{Recall that the fields in the vector hypermultiplet transform in the adjoint representation of the gauge group, and hence are neutral under the Abelian $U(1)$ gauge symmetry, while $\phi_{+}, \phi_{-}^{\dag}$ inside $\Phi$ have the same non-zero electric charge.}
\begin{align}
D_{\mu} \Phi &= \partial_{\mu} - i \mu_R \tau_3 \delta_{\mu 0} + i g A_{\mu} , \\
D_{\mu} \Psi &= \partial_{\mu} - i \mu_R \tau_3 \delta_{\mu 0} .
\end{align}
The R-charge density is
\begin{align}
Q_{R} = \Psi^{\dag} \sigma^{0} \tau_3 \Psi + i \left[ \Phi^{\dag} \left(\partial_0 - i \mu_R \tau_3 \right) \tau_3 \Phi -  \left[\left(\partial_0 - i \mu_R \tau_3\right) \tau_3 \Phi\right]^{\dag} \Phi \right] ,
\label{eq:RCharge}
\end{align}
while the electric charge density is
\begin{align}
Q_{EM} = i \left[ \Phi^{\dag} \left(\partial_0 - i \mu_R \tau_3 \right) \Phi -  \left[\left(\partial_0 - i \mu_R \tau_3\right) \Phi\right]^{\dag} \Phi \right] .
\end{align}
For future reference, note that if $\phi_{+}, \phi_{-}$ acquire identical time-independent VEVs, then $Q_R \neq 0$, while $Q_{EM} = 0$. This is the key to ensuring that a finite R-charge density does not violate the electric neutrality condition.

\subsection{Scalar ground state}
We look for time-independent scalar ground states, and work in unitary gauge, as we have done throughout the paper.  The bosonic potential with the $\mu_R$ contributions included is
\begin{align}
V^{(0)}_{\rm eff} &= \left| \sqrt{2} g a +  m \right|^2   \left(\left| \phi_+ \right|^2 + \left| \phi_- \right|^2   \right)  
+  2 g^2  \left| \phi_+ \right|^2  \left| \phi_-\right|^2 \\
&+ \frac{g^2}{2} \left(  \left| \phi_+ \right|^2 - \left| \phi_- \right|^2  -\xi^2 \right)^2
-\mu_R^2  \left( \left| \phi_+ \right|^2 + \left| \phi_- \right|^2   \right) \nonumber,
\end{align}
where $a$ is the scalar from the vector hypermultiplet.    This theory always has a stable non-trivial ground state when $\mu_R \neq 0$, which can be seen from the fact that there is no attractive $|\phi_{+}|^2|\phi_{-}|^2$ term in the potential.  Just as before,
 $a$ picks up the VEV 
\begin{align}
\langle a\rangle = -\frac{m}{\sqrt{2} g} ,
\end{align}
which is independent of $\xi$.  We will see below that charge neutrality requires that we set $\xi^2=0$, so we drop $\xi$ from here onwards.  Minimizing the scalar potential for the remaining fields we find the condition
\begin{align}
\left|\phi_+\right|^2 + \left|\phi_-\right|^2 = \frac{\mu_R^2}{g^2} .
\end{align}
To see the consequences of electric neutrality, recall that  $\phi_-^{\dagger}$ feels a chemical potential $- \mu_R$ compared to the field $\phi_+$ which feels a chemical potential $\mu_R$.  Recalling the expression for the electric charge density, it is clear that electric neutrality in the scalar sector will be ensured if they have the same VEVs,\footnote{If we had allowed $\xi \neq 0$, then the masses would of $\phi_{-}$ and $\phi_{+}$ would be split, and this argument would not work.} leading to
\begin{align}
\boxed{\left|\phi_+\right|^2 = \left|\phi_-\right|^2 = \frac{\mu_R^2}{2g^2}.}
\label{eq:RChargeVEVs}
\end{align}
Since these VEVs are non-zero for $\mu_R \neq 0$, and the scalars are charged, the $U(1)$ electromagnetic symmetry is broken, and the system is a superconductor.  Of course, the charged scalars also transform non-trivially under $U(1)_R$, so the R symmetry is also spontaneously broken once they develop VEVs.  Indeed, since both scalars develop VEVs, the R symmetry is completely broken.

\subsection{Search for a Fermi Surface}
\label{sec:FermionsR}
Paralleling the approach of the preceding sections, we again introduce a single column vector collecting all of the two-component spinors
\begin{align}
\Psi^R =\begin{pmatrix}\psi_{L \alpha} \cr \psi_{R \alpha}\cr \lambda^{\dagger \dot{\alpha}} \cr \chi^{\dagger \dot{\alpha}}  \end{pmatrix} ,
\end{align}
and rewriting Eq.~\eqref{eq:Fermions2} as
\begin{align}
\mathcal{L}_{\CN=\,2} |_{\rm fermion} &= [\Psi^{R}]^{\dag} \cdot M^{R}_{\CN =2} \cdot \Psi^{R} .
\end{align}
Now, of course, the structure of $M^R_{\CN=\,2}$ is different, since the gauginos feel the R-charge chemical potential, and the matter fermions are rendered effectively massless through the VEV of $a$, so that
\begin{align}
M^R_{\CN =2} =  \begin{pmatrix} i \bar{\sigma}^{\mu}\partial_{\mu}& 0 &i g\sqrt{2}  \phi_- & -g\sqrt{2} \phi_{+}^{\dagger}  \cr 0 & i \bar{\sigma}^{\mu}\partial_{\mu} &- ig\sqrt{2} \phi_+ &-g\sqrt{2}\phi_{-}^{\dagger}   \cr -i g\sqrt{2}  \phi_-^{\dagger} & i g\sqrt{2}  \phi_+^{\dagger} & \sigma^{\mu}\left( i \partial_{\mu} - \mu_R \delta_{\mu 0}  \right) & 0 \cr -g\sqrt{2}  \phi_+ & -g\sqrt{2}\phi_{-} &0 & \sigma^{\mu}\left( i \partial_{\mu} + \mu_R \delta_{\mu 0} \right)  \end{pmatrix}  .
\end{align}
Once we set $\phi_{+} = \phi_{-} = \phi$ in view of Eq.~\eqref{eq:RChargeVEVs}, the determinant of $M^R_{\CN=\,2}$ takes a relatively simple form.  In fact, we find it instructive to write in two different ways.  One way to write it is
\begin{align}
\det M^R_{\CN=\,2}  = &\left(\left[p_0^2 - p^2\right]\left[(p_0 + \mu_R)^2 - p^2 \right] + 8 g^2 |\phi|^2\left[p^2 - p_0( p_0 + \mu_R)\right] + 16 g^4 |\phi|^4 \right) \nonumber\\
&\times \left(\left[p_0^2 - p^2\right]\left[(p_0 - \mu_R)^2 - p^2 \right] + 8 g^2 |\phi|^2\left[p^2 - p_0( p_0 - \mu_R)\right] + 16 g^4 |\phi|^4 \right) .
\end{align}
This form makes it easy to see that the  $g^2 |\phi|^2 = 0$ consistency check is satisfied, where the determinant must reduce to one expected for four massless Weyl fermions, two without chemical potentials, and two with opposite-sign chemical potentials.  But the dispersion relations for $g^2 |\phi|^2 \neq 0$ are hard to see in this form.

The other way to write $\det M^R_{\CN=\,2}$ is
\begin{align}
\label{eq:RChargeDet}
\det M^R_{\CN=\,2}  = \prod_{i=1}^4 \left[ \left(p_0 - \tilde\mu_i \right)^2 - \left( \left|\vec p\right| + \kappa_i \right)^2 + 4 g^2 |\phi|^2   \right],
\end{align}
where
\begin{align}
\tilde\mu_{1,2} = \mu_R/2,~ \tilde\mu_{3,4} = -  \mu_R/2& &\text{and}& &\kappa_{1,3}=  \mu_R/2 ,~ \kappa_{2,4} =    - \mu_R/2 .  
\label{eq:RChargeMuKappa}
\end{align}
This makes the form of the $g^2 |\phi|^2 \neq 0$ dispersion relations for the eigenmodes manifest.  These dispersion relations are simple but quite unusual.

Setting $p_0 = 0$ to look for a Fermi surface, we find
\begin{align}
\det M^R_{\CN=\,2} |_{p_0 = 0} = \left(p^4 - p^2 \left(\mu_R^2-8 g^2 |\phi|^2\right)+16 g^4 |\phi|^4\right)^2.
\end{align}
If $g^2 |\phi|^2$ were zero, then there would be a Fermi surface at $p_F^2 = \mu^2$.  For general $g^2 |\phi|^2$, the Fermi momentum would have to satisfy the relation
\begin{align}
p_F^2 = \frac{1}{4} \left(\mu_R \pm \sqrt{\mu_R^2 - 16 g^2 |\phi|^2} \right)^2 > 0 .
\end{align}
In $\CN=\,2$ sQED, minimizing the scalar potential leads to a VEV $|\phi|^2 = \mu_R^2/(2g^2)$. As a result
\begin{align}
\det M^R_{\CN=\,2} |_{p_0 = 0} = \left(4 \mu_R^4+p^4+3 \mu_R^2 p^2\right)^2,
\end{align}
which has no real zeros.  Hence the fermions in $\CN=\,2$ sQED with a chemical potential for R-charge do not have a Fermi surface.

It is important to realize that the general structure of the fermion interaction terms in this theory is, in and of itself, compatible with the existence of Fermi surfaces, even after $U(1)$ breaking.  What prevents a Fermi surface for the fermions from appearing is the precise relationship between the normalization of the Yukawa terms and the scalar self-interaction terms, which is dictated by supersymmetry.  To see this, consider modifying the Yukawa couplings by changing $g \to g \epsilon$ and leaving everything else, including the scalar sector, unchanged.  When $\epsilon = 1$, the theory is supersymmetric, but not otherwise.  The potential Fermi momenta are then modified to
\begin{align}
p_F^2 = \frac{\mu_R^2}{4} \left(1 \pm \sqrt{1 - 8 \epsilon^2} \right)^2 > 0 .
\end{align}
Tuning $\epsilon \leq 1/(2\sqrt{2}) < 1$, a Fermi surface appears. Of course, in $\CN=\,2$ sQED, we are not allowed to vary the Yukawa couplings independently of the scalar potential, and we are stuck with $\epsilon = 1$, where there is no Fermi surface.

\section{Fermion charge density without a Fermi surface}
\label{sec:RChargeFermions}
In the preceding sections we have seen that  supersymmetric gauge theories and their cousins often do not have Fermi surfaces, despite the fact that the chemical potential couples to the fermions.  How should this result be interpreted?  Perhaps the simplest interpretation is that in the Fermi-surface-less examples all of the charge which would normally be stored by the fermions `leaks out' into the scalars through the Yukawa couplings.\footnote{We are very grateful to Julian Sonner for discussions which prodded us to explore this issue.}  In this scenario, when the fermions have no Fermi surface, the charge density would \emph{only} receive contributions from the scalar fields.  

In this section we show that this interpretation cannot be correct in general by explicitly computing the charge density $Q$  in a theory with fermions and scalars where no Fermi surface develops at finite $\mu$.  The theory we consider in this section is chosen to make the calculation of the fermion contribution to $Q$ particularly simple.  We will see that this contribution is non-vanishing.     

The general idea of the calculation is to evaluate the $T \to 0$ limit of the fermion contribution to the `grand potential' $\Omega = u - T s - \mu Q$, where $u$ is the internal energy density, $s$ is the entropy density, and $Q$ is the particle number density.  Of course, $\Omega$ also obeys the relation
\begin{align}
\Omega= - \frac{T}{V} \log Z ,
\end{align}
where $Z$ is the grand canonical partition function, $T$ is the temperature, and $V$ is the volume of the system.  Then we observe that the contributions to $\Omega$ can generically be split into a contribution from fermionic energy eigenmodes plus a contribution from bosonic energy eigenmodes, so that
\begin{align}
\Omega= -\Omega_{\rm fermions} + \Omega_{\rm bosons},
\end{align}
where the minus sign accounts for fermionic statistics when evaluating the fermion determinant in $Z$. We write $\Omega_{\rm fermions}$ and $\Omega_{\rm bosons}$ as
\begin{align}
\Omega_{\rm fermions} &=  \,\, \sum_{i \,\in\, \textrm{particles, antiparticles}}\int \frac{d^3 p}{(2\pi)^3}\, \frac{E_{p,i}}{2} + \sum_{i\,\in\,\textrm{particles}}T\int \frac{d^3 p}{(2\pi)^3}\, \log\left[1+e^{-(E_{p,i} - \mu)/T} \right] \nonumber \\
&+\sum_{i\,\in\,\textrm{antiparticles}}T\int \frac{d^3 p}{(2\pi)^3}\, \log\left[1+e^{-(E_{p,i} + \mu)/T} \right],  \\
\Omega_{\rm bosons} &=  \,\, \sum_{i \,\in\, \textrm{particles, antiparticles}}\int \frac{d^3 p}{(2\pi)^3}\, \frac{E_{p,i}}{2} + \sum_{i\,\in\,\textrm{particles}}T\int \frac{d^3 p}{(2\pi)^3}\, \log\left[1-e^{-(E_{p,i} - \mu)/T} \right] \nonumber \\
&+\sum_{i\,\in\,\textrm{antiparticles}}T\int \frac{d^3 p}{(2\pi)^3}\, \log\left[1-e^{-(E_{p,i} + \mu)/T} \right]  .
\label{eq:OmegaExpressions}
\end{align}
The dispersion relations $E_{p,i}$ one should use above are the ones appropriate to the interacting theory.  The forms above follow from a number of formalisms, with standard statistical mechanics arguments being perhaps the most physically transparent.\footnote{ Another way to obtain Eq.~\eqref{eq:OmegaExpressions} is to observe that e.g. $\Omega|_{\rm fermion} = - T \log Z|_{\rm fermion} =  -\tr \log M_D$, where $M_D$ is the appropriate Dirac operator taking into account interaction corrections to the fermion propagators, compute the trace log using one's choice of finite-T formalisms, Matsubara or Schwinger-Keldysh, and then take the $T=0$ limit.  Or one may use a $T=0$ pole prescription (which is derived from the results of the finite-T approach) to evaluate the trace log directly at $T=0$.   No matter the formalism, the result is of course the same.}  The charge density can now be defined as 
\begin{align}
Q =  -\frac{\partial \Omega}{\partial \mu} .
\end{align}
Note that the quantity $Q$ defined in this way makes sense even when symmetry associated to $\mu$ is spontaneously broken, as in the case of interest below.  (Essentially, in the condensed case, $Q V$ is the charge carried by a macroscopic lump of condensate with volume $V$.)  

We define the fermion contribution to $Q$ as 
\begin{align}
Q_{\rm fermions} =  -\frac{\partial \Omega_{\rm fermions}}{\partial \mu} .
\label{eq:QFermionDefinition}
\end{align}
So to compute $Q_{\rm fermions}$ for the theory we are interested in, we must first evaluate $\Omega_{\rm fermions}$, then take a derivative.

The theory we will focus on has two Majorana fermions $\lambda$, $\chi$, one Dirac fermion $\psi$, and one complex scalar $\phi$, with interactions defined by the Lagrangian
\begin{align}
\mathcal{L} &= \frac{1}{2}\bar{\lambda} \left(i\slashchar{\partial} + \mu \gamma_0\right) \lambda + \frac{1}{2}\bar{\chi}\left( i\slashchar{\partial} - \mu \gamma^0 \right)\chi + \bar{\psi}  i\slashchar{\partial} \psi  + |\left(\partial_{\mu}+ i\mu \delta_{\mu,0}\right)\phi|^2 \nonumber \\
&+  i g \epsilon \left( \phi^{\dagger}\bar{\psi} P_{-} \lambda - \phi^{\dagger} \bar{\lambda} P_{-} \psi - \phi \bar{\lambda} P_{+} \psi   + \phi \bar{\psi} P_{+} \lambda    \right) \nonumber \\
&-   g \epsilon \left( \phi \bar{\psi} P_{-} \chi + \phi \bar{\chi} P_{-} \psi + \phi^{\dagger} \bar{\chi} P_+ \psi + \phi^{\dagger} \bar{\psi} P_{+}\chi     \right)  
-   \frac{g^2}{2} |\phi|^4 + \mathcal{L}_{\rm CT} ,
\label{eq:ToyTheory} 
\end{align}
where $g$ and $\epsilon$ are dimensionless parameters characterizing the relative strengths of the scalar self-interactions versus the Yukawa interactions, while $\mu$ is a chemical potential for a $U(1)$ symmetry acting as $\phi \to e^{+i \alpha} \phi, \lambda \to e^{ - i\alpha} \lambda, \chi \to e^{+i \alpha} \chi$.  Finally, $\mathcal{L}_{\rm CT}$ collects the counter-terms necessary to renormalize the theory
\begin{align}
\mathcal{L}_{\rm CT} = (\delta\Lambda_{cc})^4 + (\delta m)^2 |\phi|^2+ \ldots,
\label{eq:CounterTerms}
\end{align}
and we have written only the vacuum energy $(\delta\Lambda_{cc})$ and scalar mass $(\delta m)^2$ counter-terms explicitly since it turns out that they are the only ones we will need to compute $Q_{\rm fermions}$ to the order to which we work. 

Our choice of the theory described by Eq.~\eqref{eq:ToyTheory} is inspired by $\mathcal{N}=2$ super-QED with a single matter hypermultiplet with mass $m$ and a $U(1)_{R}$ chemical potential $\mu_{R}$.  Specifically, the version of Eq.~\eqref{eq:ToyTheory} with $\epsilon = 1$ can be obtained from the $\mathcal{N}=2$ theory  by the relations $A_{\mu}=0$, $\phi_{+} = \phi_{-} = \phi/\sqrt{2}$, $a = -m/(g\sqrt{2})$, and $\mu_R = \mu$.   For our purposes in this section, the case $\epsilon = 1/\sqrt{2}$ will turn out to be the easiest to analyze. From the discussion at the end of Section~\ref{sec:FermionsR}, it follows that the fermions in the theory we consider in this section have no Fermi surface so long as $\epsilon > 1/(2 \sqrt{2})$, and this is the regime we focus on in this section.

Before looking at the interesting examples of what happens when $\epsilon > 1/(2 \sqrt{2})$, we quickly review the textbook calculation of the charge density $Q$ carried by a non-interacting Dirac fermion with a chemical potential $\mu$, which help us stay oriented during calculations in the interacting theory, which work out in an unusual way.  Following the discussion above, we write
\begin{equation}
\begin{aligned}
-\Omega(T, \mu)_{\rm Dirac} = & \,\, 4\int \frac{d^3 p}{(2\pi)^3}\, \frac{E_p}{2} + 2T\int \frac{d^3 p}{(2\pi)^3}\, \log\left[1+e^{-(E_p - \mu)/T} \right] \\
&+2T\int \frac{d^3 p}{(2\pi)^3}\, \log\left[1+e^{-(E_p + \mu)/T} \right] ,
\end{aligned}
\end{equation}
where $E_p = \sqrt{p^2 +m^2}$ is the free-fermion dispersion relation. The first term is known as the `vacuum' contribution, while the second two terms are the `matter' and `anti-matter' contributions respectively.  The factor of $4$ on the vacuum term counts the total number of degrees of freedom (spin up and spin down particle and anti-particle modes), and the factors of $2$ on the matter terms have the same origin, accounting for the spin up and down contributions.   In the zero-temperature limit, and with $\mu>0$, this reduces to
\begin{align}
-\Omega(\mu)_{\rm Dirac} =  4\int \frac{d^3 p}{(2\pi)^3}\, \frac{E_p}{2} + 2\int \frac{d^3 p}{(2\pi)^3}\, (\mu - E_p) \theta(\mu-E_p) ,
\label{eq:DiracOmega}
\end{align}
where $\theta$ is the Heaviside step function, and $\theta(\mu - E_p) = \theta(p_F)$. Of course, the anti-fermion contribution has dropped out at $T=0$. 

For the free Dirac fermion, the `vacuum' term is obviously independent of $\mu$, and is irrelevant for the charge density.  Setting $m=0$ for simplicity and evaluating the remaining `matter' term we obtain
\begin{align}
-\Omega(\mu)_{\rm Dirac} = \frac{\mu^4}{12\pi^2} \;\; \Rightarrow \;\; Q_{\rm Dirac} = \frac{\mu^3}{3\pi^2},
\label{eq:DiracCharge}
\end{align}
which is the standard result \cite{Kapusta:2006pm}.

We now turn to the calculation of the fermion contribution to $Q$ in the toy theory described by Eq.~\eqref{eq:ToyTheory}.  From Eq.~\eqref{eq:RChargeDet} and Eq.~\eqref{eq:RChargeMuKappa}, we see that we have four eigen-modes contributing to $\Omega$, with
\begin{align}
E^2_{p,i} \equiv  \left( \left|\vec p\right| + \kappa_i \right)^2 + 2 \epsilon^2 g^2 |\phi|^2 , \quad i = 1,2,3,4,
\label{eq:RChargeDispersionRelations}
\end{align}
with the $i$-th mode having the chemical potential $\tilde{\mu}_i$, but now we have $|\langle \phi \rangle|^2 = 
\mu^2/g^2$.\footnote{The normalization of $\phi$ used in this section differs from the one used in Section~\ref{sec:Rcharge}, with $\phi_{\rm here} = \phi_{\rm there}/2$, so that the kinetic term of $\phi_{\rm here}$ in Eq.~\eqref{eq:ToyTheory} is canonically-normalized. }  Note that $\tilde{\mu}_{i}$ with $i=1,2$ are positive, while $\tilde{\mu}_{i}$ with $i=3,4$ are negative for $\mu >0$.   Also, we observe that Eq.~\eqref{eq:RChargeDispersionRelations} describes eight fermionic degrees of freedom, since we have four Weyl fermions coupled to each other when $\epsilon \neq 0$.  

These dispersion relations are highly unusual, and are a consequence of the spontaneous $U(1)$ breaking driven by scalar condensation communicated to the fermions through the Yukawa couplings with strength set by $\epsilon$.\footnote{It may also be interesting to explore why happens if the $U(1)$ symmetry is broken both spontaneously and explicitly, by $U(1)$-violating mass terms.  However, the dispersion relations become very complicated in this case, and the integrals determining the fermion contribution to grand potential $\Omega$ appear to become analytically intractable.  We leave an exploration of combined spontaneous and explicit $U(1)$ breaking effects to future work.}  Hence in addition to exploring the behavior of the $\epsilon = 1/\sqrt{2}$ theory, we also verify that the $\epsilon \to 0$ limit yields the expected free-fermion results.

We now write down the fermionic contribution to $\Omega$, working with general $\epsilon$ for the moment.  Note that in view of the signs on the $\tilde{\mu}_i$'s, when writing down the matter contributions to $\Omega$ at $T=0$ we must  take into account the particle contributions for the first two modes, while for the second two modes we have to take into account the antiparticle contributions.  Adding up the contributions, we get 
\begin{align}
-\Omega(\mu)|_{\rm fermion} &=  \sum_{i=1}^{4} 2 \times \int \frac{d^3 p}{(2\pi)^3}\, \frac{E_{p,i}}{2}  
+\sum_{i=1}^{2}\int \frac{d^3 p}{(2\pi)^3}\, \left(\tilde{\mu}_i - E_{p,i}\right) \, \theta(\tilde{\mu}_i - E_{p,i}) \\
&+ \sum_{i=3}^{4}\int \frac{d^3 p}{(2\pi)^3}\, \left(-\tilde{\mu}_i - E_{p,i}\right) \, \theta(-\tilde{\mu}_i - E_{p,i})  \nonumber.
\label{eq:RChargeOmega}
\end{align}

We begin by making sure that the $\epsilon \to 0$ limit of $- \Omega|_{\rm fermions}$ behaves as expected in view of the fact that at $\epsilon=0$ no spontaneous $U(1)$ breaking is communicated to the fermions.   In the $\epsilon \to 0$, we know that the fermionic part of the theory described by Eq.~\eqref{eq:ToyTheory} becomes a theory of a single free massless Dirac fermion that feels a chemical potential $\mu$, and two free Weyl fermions which do not feel the chemical potential.  So as $\epsilon \to 0$, we must recover get Eq.~\eqref{eq:DiracCharge}.  As already noted in Section~\ref{sec:FermionsR}, the dispersion relations in Eq.~\eqref{eq:RChargeDispersionRelations} behave in a very peculiar way in this limit, so the way the consistency check is satisfied is surprisingly subtle.  Evaluating Eq.~\eqref{eq:RChargeOmega} and taking the $\epsilon \to 0$ limit, and canceling the standard UV-divergent vacuum energy contribution by adjusting the $\delta \Lambda_{cc}$ counter-term, we find that
\begin{align}
-\Omega(\mu)_{\rm fermion} =  \left(\frac{\mu^4}{96 \pi^2} + 2\times\frac{7\mu^4}{192 \pi^2} \right) = \frac{\mu^4}{12\pi^2},
\end{align}
which matches Eq.~\eqref{eq:DiracCharge} as advertised.  The unusual thing is that the first piece above comes from the \emph{vacuum} term, while the second comes from the matter and anti-matter terms.  The fact that the vacuum term makes a $\mu$-dependent contribution to $\Omega$ may seem strange, but it is a consequence of the peculiar way we must write the dispersion relations at $\epsilon = 0$ to keep them diagonal when $\epsilon > 0$. 

Now consider the same calculation when $\epsilon > 1/(2 \sqrt{2})$.  The `matter' terms in Eq.~\eqref{eq:RChargeOmega} vanish, which is the expected signature of the lack of a Fermi surface.  The `vacuum' contributions have UV divergences, as is usually the case, and must be regulated and renormalized.  For our purposes here, a simple momentum cutoff regulator $\Lambda$ is sufficient, since we are considering a Yukawa theory, see Eq.~\eqref{eq:ToyTheory}, which is a classic case where cutoff regularization is particularly efficient.%
\footnote{Dimensional regularization (DR) is also often an efficient regulator.  However, the highly unusual Lorentz-breaking dispersion relations that result after symmetry breaking make the standard DR formulas inapplicable.  Rather than common Gamma functions the analytically-continued integrals have to be written in terms of Appell functions (hypergeometric functions in two variables) in DR. However, the necessary asymptotic expansions of these functions are extremely complicated.  We were able to find verifiably-correct expressions for the relevant asymptotics in some recent mathematics literature \cite{ferreira2004asymptotic}, but these expressions are very cumbersome to work with.  It may be an interesting project in mathematical physics to figure out a way to obtain sufficiently transparent closed-form expressions for the asymptotics of these special functions for using them in QFT calculations, but it is beyond the scope of the present work.  In any case, it is a standard principle of quantum field theory that if one obtains a finite and cut-off independent expression for an observable, using a systematic regularisation and renormalisation procedure,  any other regulator would give the same final expression.  However, at intermediate stages the calculation may become more or less difficult depending on the regulator chosen, and so in the interests of simplicity we stick to cutoff regularization.}   %
We obtain
\begin{align}
-\Omega_{\rm fermion} &= \frac{\Lambda^4}{2\pi^2} +\frac{\epsilon^2 |\phi|^2 \Lambda^2}{\pi^2} \nonumber\\
&+ \left[\frac{\mu^4}{96\pi^2} + \frac{1}{2\pi^2} \epsilon^2 |\phi|^2 \left(\frac{1}{2}\epsilon^2 |\phi|^2-\mu^2\right)-\frac{1}{2\pi^2} \epsilon^2 |\phi|^2 \log (2) \left(2 \epsilon^2 |\phi|^2-\mu_R^2\right)\right.\\
&\left.+ \frac{1}{4\pi^2} \epsilon^2 |\phi|^2 \left(2 \epsilon^2 |\phi|^2-\mu_R^2\right) \log \left(\frac{8 \epsilon^2 |\phi|^2}{\Lambda ^2}\right)\right] . \nonumber
\end{align}
The power-law divergences above (together with any other ones coming from the non-fermion parts of $\Omega$) are trivially cancelled off by appropriate cosmological constant and scalar mass counter-terms from Eq.~\eqref{eq:CounterTerms}.  For generic $\epsilon$, one also has to turn on $|\phi|^4$ counter-terms at this order, and this would lead to the need to renormalize $g$ to compute $Q_{\rm fermions}$.   

However, if we consider a theory with $\epsilon = 1/\sqrt{2}$, then on the one hand there is still no Fermi surface since $1/\sqrt{2} >1/(2 \sqrt{2})$.  On the other hand, at the order to which we work above there are no logarithmic divergences proportional to $|\phi|^2$ or $|\phi|^4$.  Hence  in the theory with $\epsilon=1/\sqrt{2}$ we do not need to introduce a $|\phi|^4$ counter-term and renormalize $g$ to compute $Q_{\rm fermions}$ to leading order.  Since consideration of the theory described by Eq.~\eqref{eq:ToyTheory} with $\epsilon=1/\sqrt{2}$ is sufficient to make our point, we set 
\begin{align}
\boxed{\epsilon=1/\sqrt{2}}
\end{align}
 from here onward.  

We are now in a position to write down the renormalized expression for $\Omega_{\rm fermion}$:
\begin{align}
-\Omega_{\rm fermion}|^{\epsilon = 1/\sqrt{2}} &= -\frac{17 \mu ^4}{96 \pi ^2} \;\; \Rightarrow \;\;  \boxed{Q_{\rm fermion}|^{\epsilon = 1/\sqrt{2}} = -\frac{17 \mu^3}{24\pi^2} \neq 0 .}
\end{align}
Note that this has the same parametric dependence on $\mu$ as Eq.~\eqref{eq:DiracCharge}, but a different numerical coefficient. Looking back at Eq.~\eqref{eq:RCharge} for the total $U(1)_R$ charge, we see that in the $\epsilon = 1/\sqrt{2}$ theory it is
\begin{align}
Q|^{\epsilon = 1/\sqrt{2}} = \frac{2 \mu^3}{g^2}-\frac{17 \mu^3}{24\pi^2}+ \ldots ,
\end{align}
where the first term is the tree-level scalar contribution from the scalars, the second is the leading fermion contribution,\footnote{As usual, the fermion contribution comes from a one-loop calculation, just as in the free case: fermions are intrinsically quantum objects.} and the ellipsis denotes the one-loop scalar contribution and higher order terms. This example shows that fermions can contribute to a charge density $Q$, as defined by Eq.~\eqref{eq:QFermionDefinition}, even when there is no Fermi surface.  We emphasize that this unusual result is obtained in the unusual situation where the $U(1)$ symmetry associated to $Q$ is spontaneously broken due to scalar condensation.  For this reason, there is no conflict with Luttinger's theorem, which relates $Q_{\rm fermions}$ to the volume of the Fermi surface, because Luttinger's theorem assumes that the $U(1)$ symmetry is \emph{not} spontaneously broken.

Before closing this section, we find it illuminating to discuss how our results would be modified in a theory with a more complicated mass matrix. In any free theory with fermion-number symmetry preserving Dirac masses, the mass matrix can always be diagonalised by a linear transformation of fields with the same charge under the symmetries of the theory. After this procedure, the system is equivalent to one with free massive Dirac fermions that feel a chemical potential. The dispersion relation for the mode $i$, with mass $m_i$ that experiences a chemical potential $\mu_i$ is then given by
\begin{align}
\left(E_{i} - \mu_i \right)^2 = m_i^2 . 
\end{align}
Consequently, the charge of the system is necessarily stored in Fermi surfaces, which would appear if there are modes with $\mu_i > m_i$.  The same statement would apply in any weakly-interacting system in which the interactions do not produce effective mass terms which break the fermion number symmetries.   Such systems satisfy the assumptions that go into Luttinger's theorem, and their behavior will necessarily follow its predictions.   Symmetry preserving masses can never lead to dispersion relations of the form of Eq. \eqref{eq:RChargeDispersionRelations}, and in particular the term $g^2\phi^2$ in Eq. \eqref{eq:RChargeDispersionRelations} cannot simply be replaced by $m_{D}^2$, where $m_D$ is a Dirac mass.

The `mass terms' that arise as a result of a scalar VEV in the Lagrangian \eqref{eq:ToyTheory} spontaneously break the U(1) R-symmetry.  For example, the term $i g \epsilon \left<\phi^{\dagger}\right> \bar{\psi} P_- \lambda$ couples (a component of) the state $\lambda$, which is charged under the symmetry, to $\psi$, which is uncharged. The only way to write down a mass term which appears in the non-standard dispersion relations in the same way as $g^2|\phi|^2$ does, without spontaneous symmetry breaking, is through \emph{explicit} symmetry breaking. Such a mass means that the mass matrix cannot be diagonalised by a rotation of fields with the same charge, as opposed to in theories containing only Dirac masses.  This is not surprising, in light of the fact that such terms in the dispersion relations break the assumptions going into Luttinger's theorem. Such mass terms may arise from symmetry-breaking Majorana mass terms, which would be an explicit rather than spontaneous breaking of the symmetry. 
 
A potentially interesting calculation would be to find the charge stored in a system, qualitatively different from $\mathcal{N}=2$ theories, which contain \emph{both} symmetry-preserving Dirac and symmetry-violating masses from spontaneous symmetry breaking by a scalar (or alternatively, symmetry-breaking Majorana masses). However, the dispersion relations in such systems are extremely complicated, and even in cases where closed forms for these can be obtained, the integrals to evaluate the grand potential become very cumbersome. It would be interesting to return to this problem in future, particularly in simple non-supersymmetric theories where the dispersion relations may be tractable.

\section{Discussion}
\label{sec:Discussion}
The most familiar finite density low-temperature systems that involve chemical potentials coupling to fermions are Fermi liquids. The applicability of Landau's Fermi liquid theory requires two basic features:
\begin{enumerate}
\item A Fermi surface, showing up as {\it e. g.} the locus in spatial momentum space where the inverse fermion propagator vanishes when $p_0 = 0$.
\item Having short-ranged-enough interactions amongst its degrees of freedom. 
\end{enumerate}
These two properties lead to the existence of well-defined quasiparticles and all of the familiar Fermi liquid phenomenology like Landau's zero sound, a specific heat linear in temperature, and so on. Examples of theories which do not fit into this paradigm are intrinsically interesting, and come about when one or both of these properties fail to hold.  

Obviously, free systems satisfy both assumptions.  Perhaps the simplest non-trivial example of a non-Fermi liquid, which also happens to be relevant to our paper, is the non-supersymmetric electron plasma described by QED, which satisfies (1), but does not satisfy (2), as reviewed in Sec.~\ref{sec:Challenges}.  When there are strong attractive interactions among the fermions, one can also easily imagine (1) failing due to the formation of bosonic bound states.  The bosonic states obviously do not have a Fermi surface, and at low temperature would typically tend to Bose condense instead. If there are only parametrically weak attractive interactions between the fermions, then while the fermion Green's function will have a sharp Fermi surface singularity at any finite order in perturbation theory, the BCS mechanism generally leads to the formation of Cooper pairs and a non-perturbative BCS gap $\Delta \sim \mu e^{-1/g} \ll \mu \sim p_F$.  The Fermi surface then gets smeared out by a non-perturbatively small amount $\Delta/p_F \ll 1$.    Systems showing both sorts of behavior are well known, and have been explored in {\it e.g.}  the context of the so-called BCS-BEC crossover in cold atomic gases \cite{Chen20051,2005PhRvB..72b4534P}.   Note that in both of these examples the $U(1)$ particle number symmetry of the fermions becomes broken by \emph{composite} scalar condensation.  Luttinger's theorem does not apply once this happens.

It is much less obvious to see how a Fermi surface could disappear in perturbation theory, in the limit of of arbitrarily weak interactions, where one does not expect the fermions to be able to form bosonic bound states.  Indeed, so long as Luttinger's theorem is applicable, such a thing should not happen.  But 4D supersymmetric theories always have elementary scalar fields which couple to fermions, and these could condense even at arbitrarily weak coupling.  So for weakly coupled supersymmetric theories there is reason to be concerned about the existence of Fermi surfaces.   Our results indicate that at least some theories with interactions of the types found in supersymmetric gauge theories fail to satisfy (1) due to scalar condensation driving quadratic mixing between Dirac fermions, which directly feel the chemical potential $\mu$, and Majorana fermions, which do not. Luttinger's theorem does not apply because of scalar condensation which breaks the relevant $U(1)$ symmetry.  There does not appear to be any modified Luttinger relation of sort explored in \cite{2005PhRvB..72b4534P} that one could define in supersymmetric QED, because of the lack of separate fermionic and bosonic number symmetries.   

Moreover, as explained in Section~\ref{sec:Challenges}, in supersymmetric QED, this mixing is order unity even when the gauge coupling is arbitrarily small.  In a sharp contrast with the other examples in which  Fermi surfaces are endangered by interactions, in supersymmetric QED there is \emph{no} parameter which we could tune smoothly to interpolate between a regime where there is a perturbative Fermi surface to one where there is not.  The physics at any $g>0$ is sharply different from the physics at $g = 0$.  After the diagonalization which takes into account the scalar-condensate-induced mixing, the fermionic eigenmodes have highly peculiar dispersion relations with a complex dependence on $\mu$, and when the smoke clears we do not see a Fermi surface in any of our supersymmetric examples.  In our non-supersymmetric examples, with hard and soft breaking of SUSY, where Luttinger's theorem also does not apply, whether a Fermi surface appears depends on the values of the parameters. Perhaps this should not be surprising: just because Luttinger's theorem is not available to shield the Fermi surface from danger, this does not imply that interactions must destroy the Fermi surface. This is illustrated by our non-supersymmetric examples in Section \ref{sec:N1HardBreaking} and part of Section \ref{sec:FermionsR}, where the relevant $U(1)$ is broken, but there is nevertheless a Fermi surface. But in our supersymmetric examples, it {\emph does} turn out to be the case that turning on any non-zero interaction, which results in the $U(1)$ breaking, obliterates the Fermi surface. Finally, we again emphasize that our supersymmetric examples all led to superconducting ground states, with the $U(1)$ breaking driven by charged \emph{elementary} scalar condensation, as opposed to any sort of BCS-like fermion pairing mechanism.

Obviously, in this paper we have only managed to scratch the surface of a large pile of interesting issues.  Sticking with super-QED, or the sort of non-supersymmetric theories we considered in this paper, one can ask many questions.   For instance, what is the quasiparticle spectrum of such theories?  What are the thermodynamics? Perhaps the most conceptually interesting question is whether the fermions manage to store any of the charge density, despite not having a Fermi surface.  Relatedly, can one develop a useful heuristic understanding of the reason for the disappearance of the Fermi surface? Naively, it may have seemed that the most natural possibility is that when there is no Fermi surface, all of the charge `leaks out' of the fermion sector through the Yukawa terms, and gets stored by the scalars.  But in Section~\ref{sec:RChargeFermions}, we explicitly calculated the fermion contribution to the charge density in an example where there is no Fermi surface, and showed that the fermion contribution to the charge density is non-vanishing.  We do not yet know a heuristic physical interpretation for this result, which seems to go against the conventional wisdom about how fermions behave at finite density. Of course, this conventional wisdom is based on Luttinger-theorem-inspired pictures, and as we have emphasized Luttinger's theorem does not apply to our condensed-scalar examples.

If one bravely hopes to try to make direct contact with condensed matter physics, it may perhaps be prudent to start by attacking the questions we raised above in Abelian gauge theories, since examples of dynamical Abelian gauge fields coupled to fundamental and emergent matter of various statistics are ubiquitous in condensed matter.  Perhaps there are condensed matter systems for which theoretical models involving Yukawa interactions of the sort seen in SUSY gauge theories may be useful.   

To make contact with the results of gauge-gravity duality, it is important to generalize our analysis to include non-Abelian gauge fields, and to begin working with theories that actually have gravity duals at strong coupling.  The details of the scalar stabilization mechanisms may well be different, and presumably do not involve turning on FI terms (but see \cite{Ammon:2008va}), as we had to do here in a number of examples.   An interesting issue is that from the weak-coupling side, it seems likely that finite density would drive squark condensation, but this would lead to gauge symmetry breaking, which has not been seen in most systems at strong coupling.   (Of course, signs of breaking of \emph{global} symmetries are ubiquitous in gauge-gravity duality.) Also, instead of electrical neutrality, in weak-coupling treatments of non-Abelian gauge theories presumably a starring role would be played by color neutrality, as has been the case in studies of high density QCD.  Once the generalization to non-Abelian theories is performed, one would have the opportunity to investigate many interesting phenomenological and conceptual questions.  Is the charge typically stored in fermions, or in the scalars?  The possibility that in some cases it may be stored in scalar condensates has been noted in the AdS/CFT context in e.~g. \cite{Karch:2007br,Ammon:2011hz}.  Are there actually Fermi surfaces at weak coupling in theories that do not seem to have one holographically?  Are there examples of theories with the opposite behavior --- Fermi-surface like singularities at strong coupling, but no Fermi surfaces at weak coupling? 

We hope to return to some of these questions in future work.

\acknowledgments
\label{sec:Acknowledgements}
We are deeply indebted to Andy O'Bannon and Julian Sonner for collaboration at the initial stages of this work, and are very grateful to them for detailed and thoughtful comments on the manuscript.  We also thank Paulo Bedaque, Tom Cohen, Jason Evans, Sean Hartnoll, Nabil Iqbal, Peter Koroteev,  John March-Russell, Janos Polonyi, Subir Sachdev, Fidel Schaposnik, Andrei Starinets, and David Tong for inspirational discussions.  We are especially grateful to David Tong for his prescient skepticism regarding our original naive assumption that Fermi surfaces would surely survive the addition of SUSY interactions.   We acknowledge support from U.S. DOE grant
 FG02-94ER40823 (A.~C.),  the Graduate Scholarship of St. John's College, Oxford (S.~G.), and STFC (E.~H.).


\bibliographystyle{apsrev4-1}
\bibliography{ZeroSound}

\end{document}